\documentclass[11pt]{article}
\usepackage{amsmath,amssymb,bm,epsf,epsfig,graphicx}
\input epsf.sty
\usepackage[utf8]{inputenc}
\usepackage[english]{babel}
\usepackage{amsmath}
\usepackage{latexsym}
\usepackage{amsfonts}
\usepackage{mathtools}
\usepackage{amssymb}
\usepackage{scalerel}
\usepackage{extpfeil}
\usepackage[left=3cm,right=3cm,top=3.1cm,bottom=3.1cm]{geometry}
\usepackage{cite}
\usepackage{tensor}
\usepackage{tikz}
\usepackage{tikz-cd}
\usepackage{bbm}

\usepackage[mathscr]{euscript}
\usetikzlibrary{arrows.meta, bending}
\tikzset{C/.style={circle, minimum size=8mm,
		node contents={},
		append after command={\pgfextra{%
				\draw[-{Straight Barb[flex']}](\tikzlastnode.150) arc (450:110:2.8mm);}
	}}
}

\usepackage{hyperref}
\numberwithin{equation}{section}

    \def\bra#1{\langle #1 |}
    \def\ket#1{|#1 \rangle}

    \def\p{\partial}

    \def \be {\begin{eqnarray}}
    \def \ee {\end{eqnarray}}
    \def \bal {\begin{align}}
    \def \eal {\end{align}}
    \def \bdm {\begin{displaymath}}
    \def \edm {\end{displaymath}}

    \def\del {\partial}
    \def\0{\nonumber}

    \def\w{\omega}

\begin{document}
	\begingroup\allowdisplaybreaks
\vspace*{1.1cm}
\centerline{\Large \bf Boundary  observables in string field theory}


\vspace{.3cm}

\begin{center}

{\large Klaus Kaja$^{(a)}$\footnote{Email: kaja  at gmail.com}, Carlo Maccaferri$^{(a)}$\footnote{Email: maccafer at gmail.com}, Ulisses Portugal$^{(a)}$\footnote{Email: portugal at gmail.com}  and  Jakub Vo\v{s}mera$^{(b)}$\footnote{Email: jakub.vosmera at ipht.fr} }
\vskip 1 cm
$^{(a)}${\it Dipartimento di Fisica, Universit\`a di Torino, \\INFN  Sezione di Torino \\
Via Pietro Giuria 1, I-10125 Torino, Italy}
\vskip .5 cm
\vskip .5 cm
$^{(b)}${\it Institut de Physique Théorique\\
	CNRS, CEA, Université Paris-Saclay\\
 Orme des Merisiers, Gif-sur-Yvette, 91191 CEDEX, France}

%
\end{center}

\vspace*{6.0ex}

\centerline{\bf Abstract}
\bigskip

Starting from the gauge invariant  action for free string field theory with boundary recently constructed in \cite{paper2}, we define new gauge invariant observables which are analogous to the Brown-York charges of General Relativity. Just like the Brown-York charges, our observables originate from a boundary tadpole, and are associated to isometries of the SFT gauge group around a given background. The consistency of the construction requires the equation of motion of the background to be satisfied only at the boundary and therefore these observables can also be defined for backgrounds generated by sources in the bulk.  As examples of our construction in open string field theory, we compute the flux through the boundary of constant electromagnetic field-strength solutions and the charge associated to the Coulomb solution. As a further example in closed string field theory, we characterize the infinite conserved charges associated to stringy-haired black-hole solutions in two-dimensional string theory. We also construct a generalization of these boundary observables to the full interacting string field theory. \baselineskip=16pt
\newpage
\setcounter{tocdepth}{2}
\tableofcontents

\section{Introduction}\label{sec:intro}

Recently there has been progress in the construction of free string field theory\footnote{See \cite{Sen:2024nfd, Maccaferri:2023vns} for recent reviews on the subject.} actions in the presence of a target space with boundary \cite{georg, atakan, paper1, paper2}. In particular, starting with \cite{paper1}, we have understood how to construct fully gauge-invariant bulk-boundary actions by adding to the standard bulk degrees of freedom appropriate boundary modes which restore invariance under the gauge transformations broken by the boundary \cite{paper2}. 

In this paper we use this bulk-boundary gauge-invariant action \cite{paper2} to define  observables associated to SFT classical solutions. 

The first obvious observable is the action itself.  Thanks to the addition of the boundary contributions  to the original purely-bulk free action $\frac12\w(\Psi,Q\Psi)$ (which identically vanishes for a solution $Q\Psi=0$) this action receives non-trivial contribution from the boundary which in general do not vanish on-shell. A subtle point emerges here. In general, solutions can appear in continuous families connected to the perturbative vacuum $\Psi=0$ (this is in fact always the case in the free theory), with a non-vanishing on-shell action. How then can they have a non-vanishing (continuously varying) action different from that of the perturbative vacuum and at the same time be extrema in the variation? In fact, this  happens  because the variation of the action vanishes in the bulk but receives non-vanishing contributions at the boundary. In other words, the presence of the boundary induces a boundary tadpole. This  tadpole is typically set to zero by assuming Dirichlet boundary conditions for the fluctuations. In this way the variation of the action is still zero and the solution can be considered a valid saddle point. A simple example of this mechanism arises in electromagnetism where the gauge invariant action in presence of the boundary is $S\sim\int_M F_{\mu\nu}F^{\mu\nu} $ and its variation is given by
\begin{align}
\delta S\sim -\int_M d^Dx\, \delta A_\mu \del_\nu F^{\mu\nu}+\int_{\del M }d^{D-1}y\sqrt{\gamma}\, \delta A_\mu\, n_\nu F^{\mu\nu}\,,
\end{align}
where $n_\nu$ is the normal unit vector  and $\gamma$ is the induced metric of the boundary.
From here we clearly see that Maxwell equations $\del_\nu F^{\mu\nu}=0$ correctly set to zero the bulk part of the variation. However, for a generic solution, there is a boundary tadpole $n_\nu F^{\mu\nu}{\Big|}_{\del M}$ which is proportional to the gauge-invariant field-strength. For a generic boundary $\del M$, the action is then extremized by restricting the fluctuations to vanish at the boundary (Dirichlet condition). In this way the variational principle is satisfied and at the same time the on-shell action is non-vanishing and continuously varying. 

The existence of this boundary tadpole gives rise to a class of new boundary observables. In a sense, the boundary tadpole itself contains the new observables. However, in a generic theory, the tadpole is a state in a Hilbert space and not a scalar quantity as an observable should be. In order to extract a gauge-invariant number out of it, we need to contract it with a suitable test state in an overall gauge-invariant combination. As a matter of fact, this is precisely the situation which arises in General Relativity in the presence of boundaries \cite{York:1972sj, Gibbons:1976ue}
\begin{align}
S_{GR}=\int_M d^D x\sqrt{-g}\, R+2\int_{\del M} d^{D-1}y\sqrt{-\gamma}\,K\,,
\end{align}
where the boundary tadpole is given by the Brown-York stress tensor $T^{\rm BY}_{ab}=\gamma_{ab}K-K_{ab}$ \cite{Brown-York}
\begin{align}
    \delta S_{GR}{\Big |}_\textrm{on-shell}=\int_{\del M}d^{D-1}y \sqrt{\gamma}\, \delta\gamma^{ab}\, T^{\rm BY}_{ab}\,.
\end{align}
In this scenario there is a well-established way to extract physical observables from $T^{\rm BY}_{ab}$ \cite{Brown-York}. To do this, we pick a Cauchy surface $\Sigma$ and a boundary isometry represented by a  boundary Killing vector field $\xi^a$. With these ingredients, one can assemble the Brown-York charge
\begin{align}
    q^{\rm BY}(\xi,g_*)=\int_{\del M\cap \Sigma} d^{D-2}w \,\sqrt{h}\, \xi^a\, u^b\, T^{\rm BY}_{ab}(g_*),
\end{align}
where $g_*$ is a bulk metric obeying Einstein equations and $u^b$ is the timelike normal unit vector to $\Sigma$, pointing towards the future. This charge is conserved (it does not depend on the choice of the Cauchy surface) and it associates a conserved quantity to a given solution of Einstein equations $g^{\mu\nu}_*$ and a given boundary isometry $\xi^a$ \cite{Brown-York}.

The main result of this paper is the construction of analogous boundary charges in both free and interacting string field theories.

The paper is organized as follows. In section \ref{sec:2}, we review the standard construction of Brown-York charges in General Relativity, formulating it in a way which is particularly useful for the subsequent SFT generalization. In section \ref{sec:3}, we derive the boundary charges of free SFT with boundary and we discuss their general properties. In particular, we observe that, if the space-like component of the boundary of $M$ is compact and the solution is regular everywhere inside $M$, then the associated charges vanish. 
In section \ref{sec:examples} we give explicit examples. Subsection \ref{subsec:massless_open} deals with the on-shell action and boundary charge associated to constant field-strength solutions in free OSFT. Here we observe that in this case our conserved charge computes the flux of the (everywhere constant) field-strength through the boundary. Therefore the charge vanishes if the space-like part of the boundary is compact. We then take the opportunity of giving an example of a `solution' with a non-vanishing boundary charge, the Coulomb solution, which is killed by the BRST charge $Q$ everywhere inside $M$ except at the location of a delta-function source. In this case our charge still computes the flux of the electric-field through the space-like part of the boundary which, however, does not vanish anymore but it matches with the charge carried by the delta-function source, consistently with Gauss' law. In subsection \ref{subsec:2D black-hole} we compute  the boundary observables of the `hairy' 2D black-hole solution discussed long time ago by Mukherji, Mukhi and Sen \cite{Mukherji:1991kz} in the context of non-critical $c=1$ closed-string field theory. Here we observe the existence of infinite conserved charges, associated to the infinite-dimensional closed-string cohomology at ghost-number one, which provides a non-trivial incarnation of `stringy isometries' going beyond the standard space-time interpretation.
In section \ref{sec:interact} we generalize the boundary charges to solutions of the full interacting SFT. Here again we find that our construction only requires the non-linear interacting field equations to be satisfied at the boundary, allowing for sources in the bulk.
We conclude in \ref{sec:conclusion}.
In appendix \ref{app:A} we explicitly construct the second non-trivial element of the ghost number one cohomology in the $c=1$ closed string which corresponds to the needed stringy isometry which capture the charge associated to the first non-trivial stringy hair of the 2D black hole. In appendix \ref{app:B} we collect some of the more technical computations associated to the interacting charges discussed in section \ref{sec:interact}.

\section{A recap of Brown-York charges in General Relativity}\label{sec:2}
Consider General Relativity on a manifold $M$ with a boundary $\del M$ \cite{York:1972sj, Gibbons:1976ue}
\begin{align}
S_{GR}=\int_M d^D x\sqrt{-g}\, R+2\int_{\del M} d^{D-1}y\sqrt{-\gamma}\,K\,,\label{GR-action}
\end{align}
where $y^a$ are local coordinates on the boundary $\del M$, which is parametrized by the embedding maps $x^\mu=x^\mu(y^a)$ with $$\gamma_{ab}=\del_a x^\mu \,g_{\mu\nu}\, \del_b x^\nu\equiv e_a^\mu \,g_{\mu\nu}\, e_b^\nu\,,$$
being the induced metric and $K$ the extrinsic scalar curvature of $\del M$.
This action is gauge invariant under boundary-preserving diffeomorphisms, \footnote{By endowing the boundary with a dynamical degree of freedom describing its transverse fluctuations we can restore the full invariance under bulk diffeomorphisms (see \cite{paper1} for details), however these extra transverse gauge symmetries don't have a direct role in the definition of the Brown-York charges, and in the present discussion we  identify the gauge group of the theory as the set of boundary preserving diffeomorphisms.} 
\begin{align}
\delta_{\rm gauge}\,g_{\mu\nu}=\nabla_\mu\bar\lambda_\nu+\nabla_\nu\bar\lambda_\mu\,,
\end{align}
where the tangential vector field $\bar\lambda^\mu$  obeys
$n_\mu\bar \lambda^\mu{\big |}_{\del M}=0$, being $n_\mu$ the globally defined normal unit vector to $\del M$. 

The general variation of the action is given by
\begin{align}
\delta S_{GR}=-\int_M d^D x\sqrt{-g}\, \delta g_{\mu\nu} \left(R^{\mu\nu}-\frac12 g^{\mu\nu}R\right)-\int_{\del M} d^{D-1}y\sqrt{-\gamma}\,\delta\gamma_{ab}(K^{ab}-\gamma^{ab}K)\,.
\end{align}
When we impose Einstein equations the variation vanishes except for the boundary term
\begin{align}
\delta S_{GR}{\Big |_\textrm{on-shell}}=\int_{\del M}d^{D-1}y\sqrt{-\gamma_*}\,\delta\gamma_{ab}(\gamma_*^{ab}K_*-K_*^{ab})\,,
\end{align}
where the subscript $(\cdot)_*$ on a boundary quantity means that it is computed from a bulk metric $g_*^{\mu\nu}$ which solves Einstein equations. Notice that even if we have a solution of Einstein equations, the variation of the total action is not vanishing unless we impose Dirichlet boundary conditions on the fluctuations 
\begin{align}
\delta\gamma^{ab}=e^a_\mu\, \delta g^{\mu\nu}\, e^b_\nu{\Big|_{\del M}}=0\,.
\end{align}
In this way the solution can provide a valid saddle-point for the (semi-classical) gravitational path-integral.

However, if we relax for a moment the (very strong) requirement of Dirichlet boundary conditions, the action is no more extremized, unless we focus on very peculiar solutions and very peculiar boundaries for which the Neumann-like boundary condition $K_*^{ab}-\gamma_*^{ab}K_*=0$ is satisfied, for example flat space with a flat boundary. Nonetheless there is still important information we can extract from this boundary tadpole, which is a consequence of gauge invariance.
Indeed, if we take a peculiar metric variation given by a (boundary preserving)  gauge variation \footnote{The boundary vector field $\lambda^a$ is the pull-back of the tangential vector field $\bar\lambda^\mu$, that is $$\lambda^a=\gamma^{ab} \,e^\mu_a \,\bar\lambda_\mu\,.$$ $D_a$ is the Levi-Civita boundary covariant derivative obtained from the induced metric $\gamma_{ab}$, see for example appendix A of \cite{paper1} for more details. }
\begin{align}
\delta g^{\mu\nu}=\nabla^{(\mu}\bar\lambda^{\nu)} \quad \longrightarrow \quad\delta\gamma^{ab}=D^{(a}\lambda^{b)}\,,
\end{align}
then the off-shell gauge invariance still implies
\begin{align}
0=\delta_{\rm gauge} S_{GR}{\Big |_\textrm{on-shell}}=\int_{\del M}d^{D-1}y\sqrt{-\gamma_*}\,D_{(a}\lambda_{b)}T^{ab}\,,\label{passing}
\end{align}
where we have introduced the Brown-York stress tensor \cite{Brown-York}
\begin{align}
T^{ab}\equiv \gamma_*^{ab}K_*-K_*^{ab}=T^{ab}(g_*)\,,
\end{align}
which is by definition evaluated on-shell (i.e. for a bulk metric $g_*$ obeying Einstein equations). Continuing from \eqref{passing} and realizing that to prove gauge invariance of \eqref{GR-action},  the boundary vector field $\lambda^a$ must vanish fast enough at possible asymptotic boundaries of $\del M$ (in particular the infinite past and future of $\del M$) we can integrate by parts and obtain
\begin{align}
0=\int_{\del M}d^{D-1}y\sqrt{-\gamma_*}\,\lambda_{b}D_{a}T^{ab}\quad \longrightarrow \quad D_aT^{ab}=0\,.\label{T-cons}
\end{align}
This is the conservation law of the Brown-York stress tensor, which is required by gauge-invariance and it is ultimately a consequence of the bulk Einstein equations.

Now consider a Cauchy surface $\Sigma$ dividing $M$ (and consequently $\del M$) in two ``past'' and ``future'' parts, respectively $M_{-}$ and $M_{+}$. At the same time consider a boundary isometry $\xi^a$, that is a vector field tangent to the boundary satisfying the boundary Killing equation
\begin{align}
 D^{(a}\xi^{b)}=0\,.\label{bound-isom}
\end{align}
Notice that $\xi^a$ in general does not vanish in the asymptotic regions of $\del M$ (a simple example is a constant time translation in presence of a static solution). Let now ${\Theta_{-}(x)}$ be the characteristic function of the region $M_{-}$ which ends of the Cauchy surface $\Sigma$. Consider now using in \eqref{passing} the ``illegal'' gauge parameter $$\lambda^a\quad\longrightarrow\quad \xi^a \, \Theta_{-}(x(y))\,.$$ Since this gauge parameter does not obey fall-off conditions, we do not expect \eqref{passing} to vanish anymore and, moreover, we cannot integrate $D_a$ by parts. However we can still evaluate 
\begin{align}
D^a\lambda^b\to D^a(\xi^b \, \Theta_{-}(x(y)))=\xi^b \del^a\Theta_{-}(x(y))=\xi^b u^a \,\delta_\Sigma(y)\,,
\end{align}
where the Dirac delta distribution $\delta_\Sigma(y)$ localizes on $\del M \cap \Sigma$
\begin{align}
\int_{\del M} d^{D-1}y \sqrt{-\gamma}\,  f(y) \delta_\Sigma(y)\equiv\int_{\del M \cap \Sigma} d^{D-2}w\,  \sqrt{h} \,f(y(w))\,,
\end{align}
and $u^a$ is the (timelike) normal (minus) unit vector of $\del M \cap \Sigma$ pointing towards the future. 
The $w^i$'s are local coordinates on $\del M \cap \Sigma$ and $h_{ij}(w)$ is the (positive definite, because $\Sigma$ is a space-like surface) induced metric on $\del M \cap \Sigma$. 
The quantity we have just obtained in this way is the Brown-York charge associated to the boundary isometry $\xi$ and the classical (bulk) solution $g_*^{\mu\nu}$
\begin{align}
q^{\rm BY}(\xi,g_*)\equiv\int_{\del M \cap \Sigma}d^{D-2}w\,\sqrt{h} \,\xi^a\, u^b\, T_{ab}(g_*)\,.
\end{align}
This charge is obviously invariant under boundary diffeomorphisms which do not displace $\Sigma$. Moreover it does not depend on how we choose the Cauchy surface $\Sigma$ and as such is a conserved quantity. To verify this last fundamental property it is enough to consider the difference between two $q$-charges computed for two different choices $\Sigma_1$ and $\Sigma_2$ delimiting a bounded region of $M$ which we denote ${\cal E}_{12}$. Then it is easy to see that
\begin{subequations}
\begin{align}
&q^{(1)}(\xi,g_*)-q^{(2)}(\xi,g_*)=\nonumber\\
&\hspace{2cm}=\left(\int_{\del M \cap \Sigma_1}d^{D-2}w_1\,\sqrt{h_1}\,u_1^b\,-\int_{\del M \cap \Sigma_2}d^{D-2}w_2\,\sqrt{h_2}\,u_2^b\,\right)\xi^a\,  T_{ab}(g_*)\\
&\hspace{2cm}=\int_{\del M\cap {\cal E}_{12}} d^{D-1}y \sqrt{-\gamma} D^b(\xi^a\,  T_{ab}(g_*))=0\,,\label{pass2}
\end{align}
\end{subequations}
where we have used divergence theorem together with the fundamental properties \eqref{T-cons} and \eqref{bound-isom}.

\section{Boundary observables in free SFT}\label{sec:3}

Let us now consider the gauge invariant free SFT action constructed in \cite{paper2} 
\begin{align}
S_{\rm tot}(\Psi,\chi)=\frac12\w{\Big(}\Psi,(\Theta_M Q-\delta_{\del M}\Gamma^*) \, \Psi{\Big)}+\w{\Big(}\Psi,\delta_{\del M}\Gamma^* Q\chi {\Big)}-\frac12\w{\Big(}Q\chi,\delta_{\del M}\Gamma^* Q\chi {\Big)}\,.\label{Stot-new}
\end{align}
Here $\Theta_M(x)$ is the characteristic function of $M\subset {\mathbb R}^{1,D-1}$ and $\delta_{\del M}(x)$ is the Dirac delta distribution localizing at the boundary $\del M$.  The differential operator $\Gamma^*_{\del M}\equiv {\rm bpz}(\Gamma_{\del M})$ is defined by the relation
\begin{align}
[\Theta_M,Q]=\delta_{\del M}\Gamma^*_{\del M}+\Gamma_{\del M}\delta_{\del M}\equiv B_{\del M}\,,\label{theta-Q}
\end{align}
where \footnote{As discussed in \cite{paper2}, there is the freedom to add a generic bpz-odd correction to $\Gamma_{\del M}$, not containing transverse derivatives. In this paper we stick to the bpz-even choice \eqref{Gamma*}. }
\begin{align}
    \Gamma_{\del M}^*=c\del_\perp+\frac12 \Omega_\perp=\Gamma_{\del M}\,. \label{Gamma*}
\end{align}
Here $c$ is the zero mode of the $c$ ghost (see \cite{paper2}) and $\del_\perp$ is the normal derivative to the boundary $\del M$ (pointing outside $M$). The ghost-number one operator $\Omega_\perp=\Omega_\perp^*$ does not contain space-time derivatives and it is uniquely determined by \eqref{theta-Q}, see the next section for examples depending on the chosen background. 
As discussed at length in \cite{paper2}, this action describes the (free) dynamics of a bulk string field $\Psi(x)$ (where $x\in M$) and three boundary string fields $\chi^{-1,0,1}(y)$ (where $y\in \del M$) which are assembled in $\chi(x)$ in such a way that 
\begin{align} 
\chi^{i}(y)=\del_\perp^{i+1}\chi(x){\Big|_{\del M}}\,.
\end{align}
Higher boundary normal derivatives of $\chi(x)$ decouple from the action \eqref{Stot-new}.
This action is invariant under the gauge transformations\footnote{The full gauge transformation of the boundary mode is $\delta_{\rm gauge}\chi=\Lambda+Q\Upsilon$, \cite{paper2}, however the gauge-for-gauge parameter $\Upsilon$ does not play any role in what we are discussing.}
\begin{subequations}
\begin{align}
    \delta_{\rm  gauge}\Psi&=Q\Lambda\,,\\
    \delta_{\rm  gauge}\chi&=\Lambda\,.
\end{align}
\end{subequations}
In our argument, these gauge transformations will play a  role analogous to the boundary preserving diffeomorphisms for the GR action \eqref{GR-action}. Notice however that in the present case there is no limitation on $\Lambda$ from the choice of $\del M$. 

Varying \eqref{Stot-new} and taking advantage of $QB_{\del M}Q=0$, we get
\begin{align}
    \delta S_{\rm tot}=&\,\omega(\delta\Psi,\Theta_M Q\Psi)+\nonumber\\
    &\hspace{0.5cm}-\w(\delta\Psi, \delta_{\del M}\Gamma_{\del M}^*\Psi-\delta_{\del M}\Gamma_{\del M}^*Q\chi)-\w(\delta Q\chi, \Gamma_{\del M}\delta_{\del M}\Psi+\delta_{\del M}\Gamma_{\del M}^*Q\chi)\,.
\end{align}
The total variation clearly contains a bulk part which vanishes upon the use of the  equation of motion $Q\Psi=0$ and a generally non-vanishing boundary contribution. When we focus on a bulk solution $\Psi_*$ so that $Q\Psi_*=0$  we obtain the on-shell variation\footnote{For  our derivation we don't have to extremize the action with respect to the boundary mode $\chi$ although this would be needed in other applications, for example the computation of the on-shell action. }
\begin{align}
\delta S_{\rm tot}{\Big|}_{\rm on-shell}&=-\,\w(\delta\Psi, \delta_{\del M}\Gamma_{\del M}^*\Psi_*-\delta_{\del M}\Gamma_{\del M}^*Q\chi)+\nonumber\\
&\hspace{4cm}-\w(\delta Q\chi, \Gamma_{\del M}\delta_{\del M}\Psi_*+\delta_{\del M}\Gamma_{\del M}^*Q\chi)\,.
\end{align}
This non-vanishing boundary variation is the starting point for the construction of boundary observables. To start with, let's focus on a gauge variation  which, by gauge invariance, should give vanishing result. 
\begin{subequations}
\begin{align}
0=\delta_{\rm gauge} S_{\rm tot}{\Big|}_{\rm on-shell}&=-\w(Q\Lambda, \delta_{\del M}\Gamma_{\del M}^*\Psi_*-\delta_{\del M}\Gamma_{\del M}^*Q\chi)+\nonumber\\
&\hspace{3cm}-\w(Q\Lambda, \Gamma_{\del M}\delta_{\del M}\Psi_*+\delta_{\del M}\Gamma_{\del M}^*Q\chi)\\
&=-\w(Q\Lambda,B_{\del M} \Psi_*)\,.\label{prip}
\end{align}
\end{subequations}
Notice that the $\chi$-part has disappeared and everything depends only on the bulk solution $\Psi_*$.
If we integrate by parts (assuming $\Lambda$ vanishes fast enough in the asymptotic regions of $\del M$) we get the consistency condition
\begin{align}
    Q\,B_{\del M}\Psi_*=0\,,
\end{align}
which is identically satisfied given that $[Q,B_{\del M}]=0$ and $Q\Psi_*=0$. This consistency condition is the  analog of \eqref{T-cons}.

Now we can follow the same logical steps described in the previous section to obtain a gauge invariant, conserved charge.
To start with, let's consider a Cauchy surface $\Sigma$ dividing $M$ in the `past' $M_{-}$ and the `future' $M_+$. Let then $\Theta_-(x)$ be the characteristic function of $M_-$, in complete analogy to what we previously did.

The second ingredient  is
 a BRST-closed gauge parameter $\Xi$, the SFT generalization of an  isometry, which satisfies the generalized Killing equation \cite{Mazel:2025fxj,Cho:2025coy}
\begin{align}
Q\Xi=0\,.
\end{align}
Then we replace in \eqref{prip} the `illegal' gauge parameter
\begin{align}
    \Lambda\quad \longrightarrow\quad \Theta_-\Xi \,,
\end{align}
which does not in general obey fall-off conditions in the asymptotic regions of $\del M$,
so that we cannot integrate by parts $Q$ and find zero. On the contrary, we can compute
\begin{align}
    Q\,\Theta_-\Xi = [Q,\Theta_-]\Xi\equiv -B_\Sigma\Xi\, .
\end{align}
Here we have defined the distributional operator
\begin{align}
    B_\Sigma\equiv [\Theta_-,Q]=\delta_\Sigma \Gamma^*_\Sigma +\Gamma_\Sigma\delta_\Sigma\,,
\end{align}
which is analogous to $B_{\del M}$ but localizes on $\Sigma$ instead of $\del M$.
In this way we end up with the following quantity
\begin{align}\label{eq:SFTBrownYorkCharge}
    q(\Xi,\Psi_*)\equiv \w(\Xi,B_\Sigma\,B_{\del M}\,\Psi_*)\,,
\end{align}
which is a number we can compute for a given pair $(\Psi_*,\Xi)$, after we have fixed a boundary $\del M$ and we have chosen a Cauchy surface $\Sigma$.

This quantity has the following important properties
\begin{enumerate}
    \item Although it has been derived from a bulk-boundary gauge-invariant action, it is eventually independent of the bulk-boundary action and it is computable given a bulk solution $\Psi_*$ and an `auxiliary' boundary $\del M$.
    \item It is gauge invariant.
    \begin{subequations}
  \begin{align}
       \delta_{\rm gauge}\,q(\Xi,\Psi_*)&=q(\Xi,Q\Lambda)=\w(\Xi,B_\Sigma\,B_{\del M}\,Q\Lambda)\\
       &=\w(\Xi,Q\,B_\Sigma\,B_{\del M}\,\Lambda)=\w(Q\Xi,\,B_\Sigma\,B_{\del M}\,\Lambda)=0\,.
    \end{align} 
    \end{subequations}
    \item It does not depend on the order of $B_{\Sigma}$ and $B_{\del M}$. To see this we write
    \begin{align}
        [B_{\Sigma},B_{\del M}]=[[\Theta_-,Q],[\Theta_M,Q]]=[Q,[[\Theta_-,Q],\Theta_M]]\,.
    \end{align}
    Then, calling $n^\mu$ and $u^\mu$ the normal unit vectors of $\del M$ and $\Sigma$ respectively we can compute 
    \begin{align}
       [ [\Theta_-,Q],\Theta_M]=-c\, \delta_{\del M}\,\delta_\Sigma \,\eta_{\mu\nu}\, n^\mu\,u^\nu=-c\, \delta_{\del M}\,\delta_\Sigma \, (n\cdot u)\,.
    \end{align}
    
    This already says that if $\del M$ and $\Sigma$ are orthogonal in space-time then $[B_{\Sigma},B_{\del M}]=0$. However, even when this is not the case, when we evaluate the commutator inside our observable, we find zero
    \begin{align}
        \w(\Xi,[B_{\Sigma},B_{\del M}]\,\Psi_*)=-\w(\Xi,[Q,c\, \delta_{\del M}\,\delta_\Sigma \, (n\cdot u) ]\,\Psi_*)=0\,,
    \end{align}
    by opening the commutator, integrating by parts $Q$ thanks to the compact support $\del M\cap \Sigma$ and using that by construction $Q\Xi=Q\Psi_*=0$ on $\del M\cap \Sigma$.
    \item It does not depend on the choice of $\Sigma$ and therefore it is conserved. To see this, it is sufficient to realize that given two Cauchy surfaces $\Sigma_1$ and $\Sigma_2$, then $B_{\Sigma_2}-B_{\Sigma_1}=[Q,{\cal E}_{12}]$, where ${\cal E}_{12}=\Theta_-^{(2)}-\Theta_-^{(1)}$ which corresponds to a strictly bounded region which allows to integrate by parts $Q$ without picking contribution from the asymptotic regions of $\del M$. This property is  analogous to the conservation of the symplectic form discussed  in \cite{Bernardes:2025uzg}.
     \item It is independent of transverse displacements of the boundary $\del M$, for exactly the same reason that it is independent of transverse displacements of $\Sigma$, as we have just proven.
    
    \item The above properties hold for whatever boundary $\del M$ we choose. However if  $M\cap\Sigma$  is a compact space-like region, we can integrate by parts the $Q$ contained in $B_{\del M}=[Q,\Theta_M]$. Then this gives zero upon acting $Q$ on $\Xi$ and on $\Psi_*$. Therefore our charge formally vanishes when $M\cap\Sigma$ is compact. This may look unexpected but it can be understood as follows: our observable computes a charge carried by the solution $\Psi_*$. But if $Q\Psi_*=0$ everywhere inside $M$ then there is no charge,  because there is no source for it. In fact, for the above properties 1--5 to hold it is enough that $B_{\del M} Q\Psi_*=0$, which is localized at $\del M$. In the interior of $M$ there can be `sources'  where $Q\Psi_*\neq0$. In this case, the observable will still be invariant under continuous deformation of the boundary, as long as the boundary does not cross the source (which would break the EOM at the boundary). A useful example, which we will analyze in some detail, is the solution for a constant electric field (for which our charge will give zero for any compact $M\cap\Sigma$, by the above argument) and the (vacuum shift) solution for the Coulomb field generated by a delta-function source inside $M$, for which our charge will precisely compute the electric charge of the source, providing a new incarnation of Gauss Theorem.
\end{enumerate}

\section{Examples}\label{sec:examples}

In this section we will compute  our new boundary observable \eqref{eq:SFTBrownYorkCharge} in free SFT for a choice of open and closed string examples. 

\subsection{Flux solutions}\label{subsec:massless_open}

A generic open string field at the massless level is parameterized by the physical gauge potential $A_\mu(x)$ and a scalar auxiliary field $B(x)$
\begin{equation}
    \Psi(x) = A_\mu(x)\alpha_{-1}^\mu c_1\ket{0}-i\sqrt{\frac{\alpha'}{2}}c_0B(x)\ket{0}\,.
\end{equation}
As extensively discussed in \cite{paper2}, describing the free dynamics of these excitations as described by the open SFT kinetic term, while ensuring full gauge invariance on a manifold with boundaries, requires the introduction of boundary degrees of freedom, denoted as $\chi^{(k)}$. The general gauge-invariant bulk-boundary action takes the form
\begin{equation}\label{eq:genealaction}
    S_{\rm tot}(\Psi,\chi)=\frac12\w{\Big(}\Psi,(\Theta_M Q-\delta_{\del M}\Gamma^*) \, \Psi{\Big)}+\w{\Big(}\Psi,\delta_{\del M}\Gamma^* Q\chi {\Big)}-\frac12\w{\Big(}Q\chi,\delta_{\del M}\Gamma^* Q\chi {\Big)}\,.
\end{equation}
Here we  consider a flat Minkowski boundary $\del M$ defined by $z=0$ (where $z$ is the coordinate normal to $\p M$), for which we can write \cite{paper2}
\begin{subequations}
\begin{align}
    \Gamma^* &= -\alpha'c_0 n_\mu\del^\mu - \frac{1}{2}n_\mu\Omega^\mu \,, \\
    Q &= -\alpha'c_0\del_\mu\del^\mu - \Omega^\mu\del_\mu + Q' \,, \\[1mm]
    \Omega^\mu &= i\sqrt{2\alpha'}\sum_{n\neq0}c_{-n}\alpha_{n}^\mu \,.
\end{align}
\end{subequations}
As usual, $n_\mu$ denotes the normal vector to the boundary and $Q'$ is the part of $Q$ which does not contain space-time derivatives.
The field $\chi(x)$ should be thought of as a Taylor expansion near the boundary, at $z=0$.
\begin{equation}
    \chi(x)= \frac{i}{\sqrt{2\alpha'}}\left( \chi^{(-1)}(y)+z\chi^{(0)}(y)+\frac{1}{2}z^2\chi^{(1)}(y) + \mathcal{O}(z^3)\right)\ket{0}\,.
\end{equation}
Evaluating the action for the massless string field introduced above yields \cite{paper2}
\begin{equation}\label{eq:So_lev0}
    S(A_\mu,B,\chi^{(k)}) = \frac{\alpha^\prime}{2}\int_M d^D x \,\Big(\frac{1}{2}F_{\mu\nu}F^{\mu\nu}+\mathcal{K}^2\Big)+\frac{\alpha^\prime}{2}\int_{\del M} d^{D-1} y\, \mathcal{R}^{(0)}\mathcal{R}^{(1)}\,,
\end{equation}
where the dependence on the auxiliary fields is effectively repackaged into the following gauge-invariant combinations
\begin{subequations}
\begin{align}
    \mathcal{K}&=B-\partial_\mu A^\mu\,, \label{eq:K}\\
    \mathcal{R}^{(0)}&=\chi^{(0)}-A_z\,,\label{eq:R0}\\
    \mathcal{R}^{(1)}&=\chi^{(1)}-\partial^a\partial_a\chi^{(-1)}+2\partial_a A^a-B\,\label{eq:R1}.
\end{align}
\end{subequations}
We can then classically integrate-out these auxiliary fields (which in particular means to solve the equations of motion for the boundary modes). In this way the auxiliary sector trivially decouples, and we are left with the standard Maxwell action for the $U(1)$ gauge field
\begin{equation}
    S^*(A_{\mu})=\frac{\alpha'}{4}\int_M d^Dx \, F_{\mu\nu}F^{\mu\nu}\,.\label{F^2}
\end{equation}
Notice that without boundary modes we would have obtained Maxwell action only up to undesired non-gauge-invariant boundary contributions which can be read off from \eqref{eq:So_lev0}, setting to zero the boundary modes in (\ref{eq:R0}, \ref{eq:R1}).

We are now interested in investigating a specific background configuration, namely the constant field-strength solution. This corresponds to the case in which the gauge potential is linear in the spacetime coordinates
\begin{equation}
    A_{\mu}(x)=-\frac{1}{2}F_{\mu\nu}x^{\nu}\,,
\end{equation}
which trivially satisfies the Lorentz gauge condition $\partial_\mu A^\mu = 0$ and yields a strictly constant field-strength $F_{\mu\nu}$. Since the physical on-shell configuration requires the auxiliary field to vanish ($B(x)=0$), substituting this potential into the general string field expansion yields the exact background state
\begin{equation}
    \Psi_* = -\frac{1}{2}F_{\mu\nu}x^{\nu}\alpha^\mu_{-1}c_1\ket{0}\,.\label{flux-sol}
\end{equation}
Evaluating the action on this classical solution, and assuming  the associated boundary-fields  equations of motion, effectively setting the gauge-invariant combinations  \eqref{eq:R0}, and \eqref{eq:R1} to zero, readily gives
\begin{equation}
    S^*_{\text{on-shell}} = \frac{\alpha'}{4}F^2\, \text{Vol}_M\,,
\end{equation}
where we recognize the coefficient $\frac{1}{4}F^2$ as the canonical Maxwell lagrangian density. Since this is a continuously varying on-shell action, there must be a boundary tadpole and therefore a corresponding boundary observable.


Let us then compute the boundary observable \eqref{eq:SFTBrownYorkCharge} of the solution, keeping our boundary setting where $M$ is the half-space $z\geq0$ and the boundary $\del M$ is the flat Minkowski hyperplane, $z=0$.
As far as the choice of isometry is concerned, in this specific background of critical string theory, the only possible BRST-closed gauge parameter is the $SL(2)$ vacuum, which we normalize according to the conventions in \cite{paper2}
\begin{equation}
    \Xi = \frac{i}{\sqrt{2\alpha'}}\ket{0}\,.\label{open-iso}
\end{equation}
Substituting this gauge parameter alongside the explicit mode expansion of $\Psi_*$ into the definition of the localized charge, the full integral representation over the spacetime manifold becomes
\begin{equation}
    q(\Xi,\Psi_*) = \int d^D x \, \frac{-i}{2\sqrt{2\alpha'}} F_{\mu\nu}\bra{0} (\delta_\Sigma\Gamma_\Sigma^*+\Gamma_\Sigma\delta_\Sigma)(\delta_{\partial M}\Gamma_{\partial M}^*+\Gamma_{\partial M}\delta_{\partial M}) x^\nu\alpha_{-1}^\mu c_1\ket{0}\,.
\end{equation}
To  evaluate it, we introduce the explicit mode expansion \cite{paper2} for the boundary operator $\Gamma$, along a generic  direction $n$ (where $n=z$ is the transverse coordinate to the boundary $\partial M$, and $n=t$ for the Cauchy surface $\Sigma$)
\begin{equation}
    \Gamma^*_{n} = \alpha'c_0\partial^{n} + \frac{i\sqrt{2\alpha'}}{2}\sum_{k\neq0}c_{-k}\alpha_k^{n}\,,\quad n=z,t
\end{equation}
The non-vanishing contributions to the BPZ inner product are highly constrained by the requirement of the saturation of the ghost zero modes. Since the vacuum expectation value requires the exact ghost insertion $\bra{0}c_{-1}c_0c_1\ket{0} = 1$, the cross-terms must provide exactly one $c_{-1}$ and one $c_0$.
Moreover, since the space-time dependence is only contained in the solution and not in the isometry, only the terms of the kind $\delta\del$ will contribute while the $\del\delta$ ones will vanish. Consequently, the non-zero terms arise strictly from the cross-product between the zero-mode derivative part ($\alpha'c_0\partial^n$) of one operator and the $k=1$ oscillator part ($\frac{i\sqrt{2\alpha'}}{2}c_{-1}\alpha_1^n$) of the other. When $\Gamma_{\partial M}$ and $\Gamma^*_{\partial M}$ provide the oscillator and  $\Gamma^*_{\Sigma}$ provides the derivative,
    evaluating these operators  on the solution
    \begin{equation}
        -\left( i\sqrt{2\alpha'} c_{-1} \alpha_1^z \right) \left( \alpha' c_0 \partial^t \right) x^\nu \alpha_{-1}^\mu c_1 \ket{0}\,,
    \end{equation}
    after applying the standard canonical commutators $[\alpha_1^z, \alpha_{-1}^\mu] = \eta^{z\mu}$ and $\partial^t x^\nu = \eta^{t\nu}$, this reduces to
    \begin{equation}
        -i\sqrt{2\alpha'} \alpha' \eta^{t\nu} \eta^{z\mu} (c_{-1} c_0 c_1) \ket{0}\,.
    \end{equation}
 When, on the other hand, $\Gamma_{\Sigma}$ and $\Gamma^*_{\Sigma}$ provide the oscillator and $\Gamma^*_{\partial M}$ provides the derivative we find
    \begin{equation}
       \left( i\sqrt{2\alpha'} c_{-1} \alpha_1^t \right)  \left( \alpha' c_0 \partial^z \right) x^\nu \alpha_{-1}^\mu c_1 \ket{0}\,.
    \end{equation}
    Here, $\partial^z x^\nu = \eta^{z\nu}$ and $[\alpha_1^t, \alpha_{-1}^\mu] = \eta^{t\mu}$. 
    \begin{equation}
        i\sqrt{2\alpha'} \alpha' \eta^{z\nu} \eta^{t\mu} (c_{-1} c_0 c_1) \ket{0}\,.
    \end{equation}
Summing these contributions we end up with
\begin{align}
    q(\Xi,\Psi_*) =&\, \frac{\alpha'}{2}\int dx^D F_{\mu\nu}\delta_{\Sigma}\delta_{\partial M}\left(\eta^{\mu t}\eta^{z\nu}-\eta^{\nu t}\eta^{z\mu}\right)\\
 =&\, \alpha'F^{tz}\int dx^D \delta_{\Sigma}\delta_{\partial M}= \alpha'F^{tz} \,\textrm{Vol}(\Sigma\cap\del M)\label{eq:fluxcharge}
\end{align}
which represents the flux of the electric field through the spatial hyperplane $\partial M\cap\Sigma $. 
\subsection{Coulomb solution}
It is important to realize that, if we choose $\Sigma\cap\del M$ to be a compact spacelike manifold, for example a $D-2$ sphere,  then the charge of the flux solution we have just discussed becomes zero. This can be seen by writing $B_{\del M}=[\Theta_M,Q]=\Theta_M Q-Q\Theta_M$. In the first term in the commutator there is $Q$ which kills the solution $\Psi_*$ and in the second term, upon integration by parts (which is possible thanks to the suppression at infinity given by the compactness of  $M\cap \Sigma$),  $Q$ kills $\bra0$ from the right. This is  expected and matches the fact that the flux of the constant electric field through $\Sigma\cap \del M$ is now zero, since there is no source inside $M$. 

This suggests to consider a more `singular' solution which can have a non-vanishing boundary charge for a compact $M\cap \Sigma$. The solution that we would like to consider  is the Coulomb field emanating from a point-like static electric charge at $\vec{x}=0$.
A point-like charge $e$ appears as a source term \footnote{We can think of this as the effect of a classical fundamental string ending on the reference $D$-brane.}
\begin{align}
    \ket{{\rm source}}=e\, \delta(\vec{x})\,\alpha^t_{-1}\,c_0c_1\ket0\,,
\end{align}
corresponding to a purely time-like electric density
$$
    j^\mu=(e\, \delta(\vec{x}),\vec{0})\,.
$$
This sources a vacuum-shift equation
\begin{align}
    Q\Psi_{\rm v}=\ket{{\rm source}}\,. \label{v-sol}
\end{align}
Notice that since $\delta(\vec{x})\,\alpha^t_{-1}\ket0$ is a continuous superposition of matter primaries (by writing $\delta(\vec{x})=\int d\vec{k}e^{i\vec{k}\cdot\vec{x}}$) the source is `conserved'
\begin{align}
    Q\ket{{\rm source}}=0\,.
\end{align}
From standard cohomology arguments, the state $\ket{{\rm source}}$ is  BRST-exact for $\vec{k}\neq 0$, and only $\vec{k}= 0$ is in the cohomology, providing a potential obstruction to \eqref{v-sol}. However this can be easily remedied by continuity, slightly deforming (in the complex $\vec{k}$ space) the $\vec{k}$ integration contour ($\int d\vec{k}\to \int_\epsilon d\vec{k}$) so as to avoid $\vec{k}= 0$. The solution is readily found in Siegel gauge
\begin{align}
    \Psi_{\rm v}(\epsilon)=\frac{b_0}{L_0} \ket{{\rm source}(\epsilon)}=e\int_\epsilon d\vec{k}\,\frac{e^{i\vec{k}\cdot\vec{x}}}{\vec{k}^2}\,\alpha_{-1}^tc_1\ket0\,.
\end{align}
This corresponds to a gauge field
$$A_\mu=(V(\vec{x}),\vec{0})\,,$$
where $V(\vec{x})=e\int d\vec{k}\,\frac{ e^{i\vec{k}\cdot\vec{x}}}{\vec{k}^2}$ is the Coulomb potential (whose precise $\vec{x}$ functional form depends on the number of dimensions, contrary to its universal form in momentum space).
Notice that 
\begin{align}
  Q\Psi_{\rm v}=  Q\Psi_{\rm v}(\epsilon){\Big |}_{\epsilon\to 0}=e \int_\epsilon d\vec{k}\,\frac{ e^{i\vec{k}\cdot\vec{x}}\,\vec{k}^2}{\vec{k}^2}\alpha_{-1}^tc_0c_1\ket0{\Big |}_{\epsilon\to 0}=e\, \delta(\vec{x})\,\alpha^t_{-1}\,c_0c_1\ket0=\ket{{\rm source}}\,.
\end{align}
Let us now compute the boundary observable associated to the vacuum shift solution by choosing a compact region $M$ containing the point $\vec{x}=0$, together with the only available `isometry' \eqref{open-iso}
\begin{subequations}
\begin{align}
    & q\left(\frac{i}{\sqrt{2\alpha'}}\ket0,\Psi_{\rm v}\right)=\nonumber\\
    &\hspace{0.5cm}=\frac{i}{\sqrt{2\alpha'}}\w(\ket0,B_\Sigma B_{\del M} \Psi_{\rm v})=\frac{i}{\sqrt{2\alpha'}}\w(\ket0,B_\Sigma (\Theta_M Q-Q\Theta_M) \Psi_{\rm v})\\
    &\hspace{0.5cm}=\frac{i}{\sqrt{2\alpha'}}\w(\ket0,B_\Sigma\Theta_M \ket{{\rm source}})=e\frac{i}{\sqrt{2\alpha'}}\,{(i\sqrt{2\alpha'})}\bra{0}c_{-1}\alpha_{1}^t\,\delta_\Sigma\,\Theta_M \,\delta(\vec{x})\,\alpha_{-1}^t\,c_0c_1\ket0\\
    &\hspace{0.5cm}=e\, (-\eta^{tt}) \, \int_M\,d^Dx \,\delta_\Sigma(x)\, \delta(\vec{x})=e\,.
\end{align}
\end{subequations}
So our observable is computing the electric charge by surrounding it by the compact space-like cycle $\Sigma\cap\del M$.
\subsection{Hairy black-hole in two-dimensional string theory}\label{subsec:2D black-hole}

Here we would like to compute our boundary observables for solutions of closed SFT. A particularly useful example of such solutions is available in the non-critical $c=1$ string whose matter CFT is composed by a timelike free boson $X^0=t$ with $c=1$ and a space-like linear dilaton $X^1=\varphi$ with $c=25$. This represents a two-dimensional target space where the string coupling varies as $g_s\sim e^\varphi$, so that the theory is very weakly coupled in the region $\varphi\to-\infty$. In this region, the free SFT action is a very good approximation to the full interacting theory.  In this background we now consider the infinite family of perturbative solutions constructed in  \cite{Mukherji:1991kz}
\begin{equation}
    \Psi_* = \sum_{r>0} \lambda_r e^{\frac{r+1}{2}V\varphi} P_r\bar{P}_r c_1\bar{c}_1\ket{0}+ O(\lambda^2)\,.\label{2Dsol}
\end{equation}
The exponential dressing $e^{\frac{r+1}{2}V\varphi}(z,\bar z)$ is a weight $(1-r^2,1-r^2)$ primary and $V$ is an appropriate constant proportional to the $\varphi$ background charge.  $P_r(z)\bar P_r(\bar z)$ are the zero-momentum primaries of the $X^0=t$ CFT of weight $(r^2,r^2)$, whose first two are explicitly given by 
\begin{subequations}
\begin{align}
    P_1=&\,\alpha_{-1}^t\ket{0}\,,\\
    P_2=&\,\sqrt{\frac{2}{27}}\Big(\alpha^t_{-3}\alpha^t_{-1}-\frac34(\alpha^t_{-2})^2-\frac12(\alpha^t_{-1})^4\Big)\ket0\,,\\
    &\hspace{-2mm}\vdots \nonumber
\end{align}
\end{subequations}
In particular the solution triggered by $\lambda_1$ excites the graviton and indeed it represents the 2D black hole solution studied for example in \cite{Mandal:1991tz}, as an excitation of the linear dilaton background. Here $\lambda_1$ is to be identified with the black-hole mass \cite{Mukherji:1991kz, Mandal:1991tz}. The other parameters $\lambda_{r>1}$ correspond to ghost-number two closed string cohomology involving `massive' string states and they do not have a direct description in terms of the effective 2D dilaton-gravity. As such they can be considered as `stringy' hairs of the 2D black hole. 

We now put a one-dimensional time-like boundary $\del M$ in the weakly coupled region at $\varphi=\varphi_0\to-\infty$, which extends in time from the infinite past $t\to-\infty$ to the infinite future $t\to+\infty$. With this choice of the boundary we have \cite{paper2}
\begin{subequations}
\begin{align}
\Gamma^*_{\del M}=&\,\frac{\alpha'}{2}c_0^+\del_\varphi+\Bigg(i\sqrt{\frac{\alpha'}{2}}\Bigg)\frac12\sum_{n\neq0}\left(c_{-n}\alpha^\varphi_n+\bar c_{-n}\bar\alpha^\varphi_n\right)\,,\\
\Gamma_{\del M}=&-\frac{\alpha'}{2}c_0^+\del_\varphi^*+\Bigg(i\sqrt{\frac{\alpha'}{2}}\Bigg)\frac12\sum_{n\neq0}\left(c_{-n}\alpha^\varphi_n+\bar c_{-n}\bar\alpha^\varphi_n\right)\,,
\end{align}
\end{subequations}
where $\del_\varphi^*$ is the adjoint operator of $\del_\varphi$ with respect to the inner product of the linear dilaton zero-mode structure. In particular, given the inner product
\begin{align}
\omega^{\textrm{zero-modes}}(f(\varphi),g(\varphi))=\int_{-\infty}^\infty d\varphi\, e^{-V\varphi} \,f(\varphi)\,g(\varphi)\,,
\end{align}
we would have
\begin{align}
    \omega^{\textrm{zero-modes}}(f(\varphi),\del_\varphi^*g(\varphi))=\int_{-\infty}^\infty d\varphi\, e^{-V\varphi} \,[\del_\varphi f(\varphi)]\,g(\varphi)\,.
\end{align}
When computing the boundary observable \eqref{eq:SFTBrownYorkCharge}, we need to act $B_{\del M}$ on the solution \eqref{2Dsol} and we easily see that the $\varphi$ oscillator part of $\Gamma$ and $\Gamma^*$ does not contribute
\begin{align}
B_{\del M}\Psi_*=\left(\delta_{\del M}\Gamma^*_{\del M}+\Gamma_{\del M}\delta_{\del M}\right)\Psi_*=\frac{\alpha'}{2}\left(\delta_{\del M}\del_\varphi-\del_\varphi^*\delta_{\del M}\right)c_0^+\Psi_*\,.
\end{align}
The other boundary operator which is needed for the charge \eqref{eq:SFTBrownYorkCharge} is $B_\Sigma$, which corresponds to the choice of a constant $t$ Cauchy surface. Since the solution is time independent, it does not matter at which time we place $\Sigma$. Thanks to time independence, it will only be the $t$-oscillator part which will give contribution
\begin{align}
B_{\Sigma}\quad\longrightarrow\quad \delta_\Sigma \left(i\sqrt{\frac{\alpha'}{2}} \right)\sum_{n\neq0}\left(c_{-n}\alpha^t_n+\bar c_{-n}\bar\alpha^t_n\right).
\end{align}
With these premises, let us start by analyzing the `hair-less' BH solution obtained by setting $\lambda_r = 0$ for all $r > 1$
\begin{equation}\label{eq:2DBH_state}
    \Psi_1 = \lambda_1 e^{V\varphi}\alpha_{-1}^t\bar{\alpha}_{-1}^t c_1\bar{c}_1\ket{0}\,.
\end{equation}
The only missing ingredient to compute the boundary observable is the choice of a proper isometry sensitive to $\lambda_1$,  that is a  ghost-number one element of the closed string cohomology with a non-trivial overlap with $\Psi_1$ in \eqref{eq:SFTBrownYorkCharge}. It is not difficult to realize that this is the generator of time translations (using the normalization from \cite{paper2})
\begin{equation}
    \Xi_1 = \frac{i}{2\sqrt{2\alpha'}} \big( c_1\alpha_{-1}^t - \bar{c}_1\bar{\alpha}^t_{-1} \big) \ket{0} \,.
\end{equation}
Now we have all the ingredients to compute the boundary observable
\begin{subequations}
     \label{eq:BH_charge_calculation}
\begin{align}
    q(\Xi_1, \Psi_{1}) &= i\frac{\lambda_1}{2\sqrt{2\alpha'}} \int dt d\varphi \, e^{-V\varphi} \bra{0} \big( c_{-1}\alpha_1^t - \bar{c}_{-1}\bar{\alpha}_1^t \big) c_0^- B_\Sigma B_{\del M} \alpha_{-1}^t \bar{\alpha}^t_{-1} c_1\bar{c}_{1} e^{V\varphi} \ket{0}  \\[2mm]
    &= i\frac{\lambda_1}{2\sqrt{2\alpha'}} \Bigg(i\sqrt{\frac{\alpha'}{2}}\Bigg) \frac{\alpha'}{2} \int dt d\varphi\, e^{-V\varphi} \bra{0} \big( c_{-1}\alpha_1^t - \bar{c}_{-1}\bar{\alpha}_1^t \big)\times\nonumber \\ &\hspace{1cm} \times c_0^-  \delta_\Sigma \sum_{n\neq0}\left(c_{-n}\alpha^t_n+\bar c_{-n}\bar\alpha^t_n\right)c_0^+ \big( \delta_{\partial M}\partial_\varphi - \partial_\varphi^*\delta_{\partial M} \big) \alpha_{-1}^t \bar{\alpha}^t_{-1} c_1\bar{c}_{1} e^{V\varphi} \ket{0}  \\[2mm]
    &= -\frac{\alpha'}{8} \int dt\,d\varphi \, e^{-V\varphi} \bra{0} \big( c_{-1}\alpha_1^t - \bar{c}_{-1}\bar{\alpha}_1^t \big)\times\nonumber \\ &\hspace{3cm} \times c_0^- \delta_\Sigma \left(c_{-1}\alpha^t_1+\bar c_{-1}\bar\alpha^t_1\right)c_0^+\big( \delta_{\partial M}\partial_\varphi \big) \alpha_{-1}^t \bar{\alpha}^t_{-1} c_1\bar{c}_{1} e^{V\varphi} \ket{0}  \\[2mm]
    &= \frac{\alpha'}{8} \lambda_1 V  \,.
\end{align}
\end{subequations}
Thus the boundary observable associated to time translations is directly proportional to the black hole mass parameter $\lambda_1$.

As far as the other charges $\lambda_{r>1}$ are concerned, in order to detect them, one has to use different isometries in the ghost number-one sector.
The general structure of such isometries is rather complicated and it is related to the existence of non-trivial cohomology in the (anti)holomorphic sector at ghost number zero \cite{Imbimbo:1991ia}. They can be written as \cite{Sen:2004yv}
\begin{subequations}
\begin{align}
\Xi_r=&\,(c_1P_r \bar X_r-X_r \bar c_1\bar P_r)e^{\frac{1-r}{2}V\varphi}\ket0\,,\\
Q\Xi_r=&\,0\,.
\end{align}
\end{subequations}
Here $X_r$ is a ghost-number zero operator of weight $r^2-1$ containing $t$, $\varphi$ as well as $b$ and $c$ oscillators, \cite{Imbimbo:1991ia}. 
Appropriately normalizing these isometries, we expect to find 
\begin{align}
q(\Xi_r, \Psi_*)=\lambda_r\,. \label{BH charge}
\end{align}
In Appendix \ref{app:A}, we explicitly verify this for the first non-trivial hair $\lambda_2$. 
Notice that we expect the result \eqref{BH charge} to remain true in the interacting theory since we are free to put the boundary at any $\varphi_0$ in particular at $\varphi_0\to-\infty$ where the theory is free.
\section{Boundary observables in interacting SFT}\label{sec:interact}

Let us now explore how the conserved boundary charge \eqref{eq:SFTBrownYorkCharge} generalizes to fully interacting theories endowed with an $L_\infty$ structure. As opposed to the free case, in the interacting case we do not have a consistent variational principle at our disposal. That is to say, we do not have a gauge-invariant action with boundary terms such that its variation could be set to zero by assuming equations of motion in the bulk and boundary conditions which would not overconstrain the problem. Instead of deriving the charge by means of the procedure used in Sections \ref{sec:2} and \ref{sec:3}, we will therefore proceed by simply writing down a proposal for the charge and then proving that it possesses all desired properties. As before, we will see that instead of having to assume that the charge is evaluated on a field configuration $\Psi$ which is on-shell everywhere in the bulk of $M$, we can afford to only require that $\Psi$ is on-shell on $\partial M$, that is, in all expressions containing derivatives of $\Theta_M$.
We will also continue to assume that one can integrate by parts along the intersection $\partial M\cap \Sigma$.

\subsection{Isometry of a classical solution $\Psi$}

The construction is built upon the notion of an isometry of a classical field configuration $\Psi$. Starting with $\Psi$, this is a string field which is assumed to satisfy the full equation of motion
\begin{align}
Q\Psi+\sum_{n\geq2}\frac1{n!} l_n(\Psi^{\wedge n})=0\,.\label{eq:Linfty}
\end{align}
This holds up to possible contributions from sources which, however, will be taken to vanish on $\p M$.
Here the symmetric multi-linear maps $l_n:\mathcal{H}^{\wedge n}\longrightarrow \mathcal{H}$ obey the relations of an $L_\infty$ algebra and, up to boundary terms, are cyclic with respect to the symplectic pairing $\omega$. We will assume that $\Psi$ belongs to a continuous family of classical field configurations which is usually referred to as \textit{pre-phase space} (see \cite{Bernardes:2025uzg} for a recent review of the covariant phase-space formalism). It then makes sense to consider small variations $\Psi\longrightarrow \Psi+\delta \Psi$ which preserve the equation of motion \eqref{eq:Linfty}. We will interpret the differential $\delta \Psi$ as a 1-form on the pre-phase space and therefore take it to be an anti-commuting object. Quotienting the space of classical field configurations $\Psi$ by the gauge variations, we arrive at the concept of the \textit{phase space}. Here an infinitesimal gauge variation can be geometrized in terms of the Lie derivative along a string ``vector'' field $\Lambda_\Psi$, which plays the role of the gauge parameter. In other words, the infinitesimal gauge variation reads
\begin{align}
  \mathcal{L}_{\Lambda_\Psi} \Psi=\iota_{\Lambda_\Psi} \delta \Psi =  Q\Lambda_\Psi +\sum_{n\geq 1} \frac{1}{n!}l_{n+1}(\Psi^{\wedge n},\Lambda_\Psi):= Q_\Psi \Lambda_\Psi\,.
\end{align}
In the final equality, we have introduced the operator $Q_\Psi$ which can be thought of as the kinetic operator of the SFT action for small fluctuations around the classical vacuum $\Psi$. As a consequence of $\Psi$ satisfying the equation of motion \eqref{eq:Linfty}, one can show that this ``shifted BRST operator'' is nilpotent, that is
\begin{align}
    Q_\Psi^2 =0\,.
\end{align}
Furthermore, in the particular case when the fluctuations preserve the pre-phase space, we recover the condition
\begin{align}
    Q_\Psi \delta\Psi=0\,,
\end{align}
which can be derived by varying the equation of motion \eqref{eq:Linfty}. Finally, $\iota_\Lambda \delta\Psi$ denotes the contraction (interior product) of the vector field $\Lambda_\Psi$ with the 1-form $\delta \Psi$ on pre-phase space. By virtue of the Cartan's magic formula, the action of the Lie derivative $\mathcal{L}_{\Lambda_\Psi}$ can then be generally computed as $\iota_{\Lambda_\Psi}\delta +\delta\iota_{\Lambda_\Psi}$.

An isometry of a classical field configuration $\Psi$ 
will be represented by the degree-odd string vector field $\Xi_\Psi$ (carrying the same ghost number as the gauge parameter $\Lambda_\Psi$) satisfying the generalized Killing equation
\cite{Mazel:2025fxj,Cho:2025coy}
\begin{align}
    Q_\Psi \Xi_\Psi =0\,.\label{eq:Xi}
\end{align}
A variation $\delta\Psi$ along the pre-phase space then has to be accompanied by a variation $\delta\Xi_\Psi$ of the string vector field $\Xi_\Psi$, which satisfies
\begin{align}
    Q_\Psi \delta\Xi_\Psi + l_2^\Psi(\delta\Psi,\Xi_\Psi)=0\,,\label{eq:delXi}
\end{align}
which, in turn, can be obtained by differentiating \eqref{eq:Xi}.
Here 
\begin{align}
    l_2^\Psi(\cdot,\cdot) = \sum_{n\geq 2} \frac{1}{n!} l_{n+2}(\Psi^{\wedge n}, \cdot,\cdot)
\end{align}
denotes the string 2-product shifted around the classical vacuum $\Psi$. Again, as a consequence of $\Psi$ solving the equations of motion \eqref{eq:Linfty}, one can show that $Q_\Psi$ is a graded derivation of $l_2^\Psi$, namely 
\begin{align}
    Q_\Psi l_2^\Psi(A,B) =- l_2^\Psi( Q_\Psi A,B) -(-1)^{d(A)}l_2^\Psi( A, Q_\Psi B)\,.
\end{align}
Equation \eqref{eq:delXi} can then be solved by assuming that associated to $Q_\Psi$, there is a contracting homotopy operator $h_\Psi$, which satisfies a Hodge-Kodaira decomposition
\begin{align}
  1=  Q_\Psi h_\Psi + h_\Psi Q_\Psi + P_\Psi  \label{eq:HK}
\end{align}
of the identity. Here we will take $P_\Psi$ to be the projector on cohomology of $Q_\Psi$, so that it satisfies
\begin{align}
    P_\Psi Q_\Psi = Q_\Psi P_\Psi =0\,.\label{eq:PQ}
\end{align}
On the other hand, the contracting homotopy operator is assumed to project \textit{out} of the cohomology and be itself nilpotent, namely
\begin{align}
    P_\Psi h_\Psi = h_\Psi P_\Psi =0 = h_\Psi^2\,.\label{eq:side}
\end{align}
Using this structure, we can then write the general form of $\delta\Xi_\Psi$ as
\begin{align}
    \delta\Xi_\Psi = -h_\Psi l_2^\Psi(\delta\Psi,\Xi_\Psi)+\xi_\Psi\,,\label{eq:delXisol}
\end{align}
that is at least provided that there is no (cohomological) obstruction to extending the isometry generator $\Xi_\Psi$ from $\Psi$ to $\Psi+\delta\Psi$. Such an obstruction would have been measured by the state  
\begin{align}
    P_\Psi l_2^\Psi(\delta\Psi,\Xi_\Psi)\,,\label{eq:obs}
\end{align}
which we need to assume to vanish if \eqref{eq:delXisol} is to provide a solution to \eqref{eq:delXi}. In the cases where \eqref{eq:obs} is non-zero, this can be interpreted in terms of the classical solution developing an asymmetry as one travels across the phase space.\footnote{In the special case when the 1-form \eqref{eq:obs} is contracted with a gauge parameter $\Lambda_\Psi$, one obtains 
\begin{align}
    \iota_{\Lambda_\Psi}  P_\Psi l_2^\Psi(\delta\Psi,\Xi_\Psi) = -P_\Psi l_2^\Psi(Q_\Psi\Lambda_\Psi,\Xi_\Psi)=P_\Psi Q_\Psi l_2^\Psi(\Lambda_\Psi,\Xi_\Psi)=0\,,
\end{align}
meaning that $\Xi_\Psi$ can be always extended along the gauge-trivial directions of the pre-phase space without obstruction. Up to $Q_\Psi$-exact states, the r.h.s.\ of \eqref{eq:delXisol} then reduces to the corresponding gauge transformation of $\Xi_\Psi$, namely $\mathcal{L}_{\Lambda_\Psi}\Xi_\Psi = l_2^{\Psi}(\Lambda_\Psi,\Xi_\Psi)$.}
Furthermore, $\xi_\Psi$ on the r.h.s.\ of \eqref{eq:delXisol} is an arbitrary generator of an isometry at $\Psi$, that is, a state in the $Q_\Psi$-cohomology at the corresponding ghost number. That being said, in the following, we will be interested purely in the variations of $\Xi_\Psi$ which are \textit{induced} by the variation $\delta\Psi$ along the pre-phase space and which do not change the direction in the space of independent isometries. This amounts to requiring
\begin{align}
    \xi_\Psi=0\,.
\end{align}
Finally, assuming that we fix an isometry $\Xi_0$ at the perturbative vacuum $\Psi=0$ of the theory, equation \eqref{eq:delXisol} can be integrated from $\Psi=0$ to give an implicit expression for $\Xi_\Psi$
\begin{align}
    \Xi_\Psi = \Xi_0 -\int_0^\Psi  h_{\Psi'} l_2^{\Psi'}(\delta\Psi',\Xi_{\Psi'})\,,\label{eq:Xiimplicit}
\end{align}
where $\Psi'$ denotes the integration variable. Indeed, by carefully evaluating the induced variations of $h_\Psi$ and $l_2^\Psi$, one can show\footnote{See Appendix \ref{app:B} for details.} that the integrand on the r.h.s.\ of \eqref{eq:Xiimplicit} is a closed 1-form so that the integral on the pre-phase does not depend on the choice of a path (modulo potential global issues). Introducing a coordinate system $\lambda^i$ on the pre-phase space and choosing an arbitrary path $\lambda^i(t)$ for $0\leq t\leq 1$ such that $\Psi(\lambda^i(0))=0$ and $\Psi(\lambda^i(1))=\Psi$, the integral on the r.h.s.\ of \eqref{eq:Xiimplicit} can be concretely computed as
\begin{align}
   \int_0^\Psi  h_{\Psi'} l_2^{\Psi'}(\delta\Psi',\Xi_{\Psi'})= \int_0^1 dt\, \dot{\lambda}^i(t) \, h_{\Psi(t)} l_2^{\Psi(t)}(\p_{i}\Psi,\Xi_{\Psi(t)})\,,
\end{align}
where we have abbreviated $\Psi(t)\equiv \Psi(\lambda^i(t))$.
Iterating, one can finally express the isometry $\Xi_\Psi$ at a generic point $\Psi$ in pre-phase space as a Dyson series 
\begin{align}
    \Xi_\Psi = \Xi_0 -\int_0^\Psi  h_{\Psi'} l_2^{\Psi'}(\delta\Psi',\Xi_{0})+\int_0^\Psi \int_0^{\Psi'} h_{\Psi'} l_2^{\Psi'}(\delta\Psi',h_{\Psi''} l_2^{\Psi''}(\delta\Psi'',\Xi_{0}))+\ldots\label{eq:dyson}
\end{align}
in terms of the corresponding isometry generator $\Xi_0$ at the perturbative vacuum $\Psi=0$. In other words, equation \eqref{eq:dyson} therefore defines a vector field on the pre-phase space generating a particular isometry of a classical field configuration $\Psi$ as a functional of a reference isometry $\Xi_0$ at $\Psi=0$. 

\subsection{Surface charge}

We are now in a position to write down the main ingredient in the construction of the conserved charge, namely the \textit{charge 1-form}
\begin{align}
    \eta(\Xi_\Psi,\Psi) = \omega(\Xi_\Psi, B^{\Psi}_\Sigma B^{\Psi}_{\partial M}\delta \Psi)\,.\label{eq:eta}
\end{align}
In writing down $\eta(\Xi_\Psi,\Psi)$, we have introduced the operators
\begin{subequations}
\begin{align}
    B^{\Psi}_{\partial M} &=[\Theta_M,Q_\Psi]\,,\\
    B^{\Psi}_{\Sigma} &=[\Theta_-,Q_\Psi]\,,
\end{align}
\end{subequations}
which together localize the 1-form \eqref{eq:eta} on the intersection $\partial M\cap \Sigma$. Similarly to the free case discussed in Section \ref{sec:3}, the order of the two boundary operators $B^{\Psi}_{\partial M}$ and $B^{\Psi}_{\Sigma}$ in the definition of $\eta(\Xi_\Psi,\Psi)$ does not matter. Indeed, we have
\begin{align}
    [B^{\Psi}_{\partial M},B^{\Psi}_{\Sigma}] = [C_\Psi,Q_\Psi]+[\Theta_-,[\Theta_M,Q_\Psi^2]]\,,
\end{align}
where $C_\Psi\equiv [[\Theta_M,Q_\Psi],\Theta_-]$ and the second term vanishes because the commutator with $\Theta_M$ localizes $Q_\Psi^2$ on $\p M$ where it is zero as a consequence of the equations of motion \eqref{eq:Linfty} satisfied by $\Psi$. Finally, since $C_\Psi$ localizes on $\p M \cap \Sigma$, we would have had
\begin{align}
    \omega(\Xi_\Psi,[C_\Psi,Q_\Psi]\delta \Psi)=\omega(\Xi_\Psi,C_\Psi Q_\Psi\delta \Psi)+\omega(Q_\Psi\Xi_\Psi, C_\Psi\delta \Psi)=0\,,
\end{align}
thus proving that $[B^{\Psi}_\Sigma, B^{\Psi}_{\partial M}]\delta \Psi$ would have given zero upon contraction with $\Xi_\Psi$ using $\omega$.

As we can see, $\eta(\Xi_\Psi,\Psi) $ can be thought of as a functional of the isometry vector string field $\Xi_\Psi$ and it depends on the position $\Psi$ in the pre-phase space.
Assuming that \eqref{eq:Linfty} is satisfied everywhere in the bulk of $M$, we are then able to prove the following properties of the 1-form $\eta(\Xi_\Psi,\Psi)$ :
\begin{enumerate}
    \item It is closed 
    \begin{align}
        \delta\eta(\Xi_\Psi,\Psi)=0
    \end{align}
    and therefore (locally in the pre-phase space) exact. (See Appendix \ref{app:B} for a proof.)
    \item Contracting with any string vector field $\Lambda_\Psi$ from the space of gauge parameters, we obtain
    \begin{align}
        \iota_{\Lambda_\Psi} \eta(\Xi_\Psi,\Psi) = \omega(\Xi_\Psi, B^{\Psi}_\Sigma B^{\Psi}_{\partial M}Q_\Psi\Lambda_\Psi)=0\,. 
    \end{align}
    (This is quite immediate because $Q_\Psi$ anticommutes with both boundary operators and since the expression localizes on the intersection $\p M\cap \Sigma$, it can be integrated by parts to kill $\Xi_\Psi$ on the bra.)
    \item It is gauge-invariant
    \begin{align}
        \mathcal{L}_{\Lambda_\Psi} \eta(\Xi_\Psi,\Psi)=0
    \end{align}
    (This is a direct consequence of properties 1.\ and 2.\ together with the Cartan's magic formula.)
    \item It is conserved, namely it does not depend on the choice of the Cauchy surface $\Sigma$. 
    \item It does not depend on the transverse position of $\partial M$, as long as $\p M$ does not cross any sources.
    \end{enumerate}
The proofs of properties 4.\ and 5.\ are completely parallel to the free case discussed in Section \ref{sec:3}.
By virtue of property 1., one can then measure the change in the charge associated to the isometry $\Xi_0$ at the perturbative vacuum $\Psi=0$ as we move to a generic point $\Psi$ of the pre-phase space as
\begin{align}
    q(\Xi_\Psi,\Psi) = q(\Xi_0,\Psi=0)+\int_0^\Psi \eta(\Xi_{\Psi'},\Psi')\label{eq:int_charge}
\end{align}
independently of the integration path.
At the same time, as a consequence of property 2., the integral on the r.h.s.\ of \eqref{eq:int_charge} is 0 whenever $\Psi$ is connected to $\Psi=0$ by a gauge transformation. The expression \eqref{eq:int_charge} therefore defines the charge $q(\Xi_\Psi,\Psi)$ to be a (gauge-invariant) function on the phase space for any such isometry $\Xi_0$ at $\Psi=0$ which can be extended throughout the phase space without obstruction. Moreover, properties 4.\ and 5.\ ensure that $q(\Xi_\Psi,\Psi)$ yields a conserved quantity with the surface $\p M$ entering purely topologically into its computation. Furthermore, note that we can write
\begin{align}
    \eta(\Xi_\Psi,\Psi)=\iota_{\Theta_M \Xi_\Psi}\Omega\,,
\end{align}
where $\Omega=-(1/2)\omega(\delta\Psi,B_\Sigma^\Psi\delta \Psi)$ is the BEF symplectic form \cite{Bernardes:2025uzg}.
Finally, taking $\Psi$ to be close to the perturbative vacuum and linearizing the r.h.s.\ of \eqref{eq:int_charge}, one recovers the ``free'' charge \eqref{eq:SFTBrownYorkCharge} derived above.\footnote{Also appropriately setting the ``zero-point'' of the charges by fixing $q(\Xi_0,\Psi=0)=0$.} 

In summary, these arguments constitute a convincing body of evidence for \eqref{eq:int_charge} being the correct generalization of the charge constructed in Section \ref{sec:3} to the case of an interacting theory. At the same time, since neither $\Theta_-$ nor $\Theta_M$ have to be taken as sharp step functions (and can instead be constructed as ``sigmoids'' \cite{Bernardes:2025uzg}), this charge appears to be well suited for a theory with non-local interactions such as SFT.

\section{Discussion and outlook}\label{sec:conclusion}

In this paper, we have constructed new observables in string field theory which, to a classical field configuration together with a choice of an isometry and a co-dimension one surface in spacetime, associate conserved, gauge-invariant charges.
This achievement opens up several exciting future explorations, part of which we list here.
\begin{itemize}
\item It would be interesting to test the interacting form of the observables \eqref{eq:int_charge} in the case of some known analytic solution, for example  \cite{Erler:2014eqa, Erler:2019fye, Ishibashi:2016xak}.

 \item Taking advantage of the fact that our boundary observable requires the equations of motion to be satisfied only at the boundary,  we can use it to capture  D-branes observables by computing their back-reaction on the closed string geometry in the form of a   vacuum-shift solution sourced by the boundary state of the back-reacting D-brane system, in analogy to what we did for the Coulomb solution. In this perspective it would  be interesting to relate our boundary observables to the conserved charges for asymptotic field configurations discussed in \cite{Sen:2004yv}.

\item It would be interesting to explore what kind of information these boundary observables can bring to the understanding of open/closed duality in SFT \cite{Maccaferri:2023gof}. For example we could characterize the moduli space of closed SFT solutions of the $(2,1)$ minimal string theory \cite{Gaiotto:2003yb}, following an analogous computation to what we did here for the infinite conserved charges of the two-dimensional black-hole in the $c=1$ string. These $(2,1)$ moduli could then be compared to D-brane charges extracted from the dual FZZT brane construction giving rise to the Kontsevich model.

\item It would be conceptually important to have a  gauge-invariant formulation of interacting SFT with boundary, with the necessary boundary modes to guarantee gauge invariance,  from which we could derive the interacting boundary observables \eqref{eq:int_charge}, following analogous logical steps  to what we have used in the free theory. In particular, assuming the existence of an interacting bulk-boundary theory with an appropriate homotopy structure generalizing \cite{paper2}, it would be interesting to see how the boundary modes will decouple in the construction of \eqref{eq:int_charge}.


\item On general grounds, it would be  very interesting to have a SFT construction of (quantum) black-hole entropy based on boundary observables.

\end{itemize}

We hope the present results will help in better understanding the observables of string theory and quantum gravity. 
\vspace{1cm}

{\bf Note added}:   While the writing of this paper was nearly completed, the preprint \cite{Bernardes:2026ofg} appeared, which discusses conserved charges from an Hamiltonian approach and recasts in this perspective the Brown-York charges of General Relativity. It will be interesting to clarify the relations of this approach with ours.

\section*{Acknowledgments}
We thank for stimulating discussions the participants of the workshop ``String Field Theory and Flux Compactification'' Benasque,  10-23 May 2026, where  this work was preliminarily presented.  The work of CM,  KK and UP is partially supported by the MUR PRIN contract 2020KR4KN2 ``String Theory as a bridge between Gauge Theories and Quantum Gravity'' and by the INFN project ST$\&$FI ``String Theory and Fundamental Interactions''. The work of JV is supported by the ERC Starting Grant 853507.

\appendix

\section{Higher black hole isometries} \label{app:A}

In this appendix we compute the higher level isometry of the $2d$ string which will allow us to compute the $\lambda_2$ charge of the black hole. This isometry is the a state in the BRST cohomology at ghost number $1$ and level $3$. This state can be written in terms of the holomorphic and anti-holomorphic cohomologies in the form
\begin{equation}\label{Xi_2}
    \Xi_2 = \frac{i}{\sqrt{2\alpha'}}\left(\Xi_{2L}^{(0)}\otimes\Xi_{2R}^{(1)} - \Xi_{2L}^{(1)}\otimes\Xi_{2R}^{(0)}\right)\,,
\end{equation}
where $\Xi_{2L}^{(n)}(\Xi_{2R}^{(n)})$ is a left-(right-)moving state in the ghost-number $n$ cohomology. $\Xi_L^{(1)}$ is simply given by
\begin{equation}
    \Xi_{2L}^{(1)} = c_1P_2e^{-\frac{1}{2}V\varphi}|0\rangle\,.
\end{equation}
The most difficult part is computing $\Xi^{(0)}$. The ghost number zero cohomology was computed in \cite{Imbimbo:1991ia} for $c<1$ strings. Although their construction does not work in general for determining the $c=1$ cohomology, it turns out that their result for the ghost-number zero level $3$ state is valid even in this case. Namely, we have
\begin{align}\label{Ghost number zero state}
    \Xi^{(0)}_{2L} = &\left[4 b_{-3}+2\left(L_{-1}^{(\varphi)}-L_{-1}^{(t)}\right) b_{-2} + \right. \nonumber\\
    &\left. \hspace{1cm}+\left(2 L_{-2}^{(\varphi)}-2L_{-2}^{(t)} + L_{-1}^{(\varphi)2}+L_{-1}^{(t)2}-L_{-1}^{(\varphi)} L_{-1}^{(t)}\right) b_{-1}\right]c_1\alpha^t_{-1}e^{-\frac{1}{2}V\varphi}|0\rangle \,.
\end{align}
It is straightforward to check that, acting with the BRST charge on this state, we get
\begin{equation}  \left(L_{-1}^{(\varphi)3}+4L_{-2}^{(\varphi)}L_{-1}^{(\varphi)}+6L_{-3}^{(\varphi)} - \frac{1}{2}L_{-1}^{(t)3}+2L_{-2}^{(t)}L_{-1}^{(t)}-L_{-3}^{(t)}\right)c_1\alpha^t_{-1}e^{-\frac{1}{2}V\varphi}|0\rangle\,,
\end{equation}
which vanishes upon writing the Virasoro generators in terms of $X^0$ and $\varphi$ oscillators.

To prove that $\Xi_2$ is a physical state, we should also check that it is not BRST exact. But this is guaranteed to be true if the charge \ref{eq:SFTBrownYorkCharge} computed with $\Xi_2$ is nonzero, which we will now show to be true. First, we note the $c_0^+$ necessary to saturate the ghost number in the correlator can only come from $B_{\partial M}$, so that
\begin{equation}
    \w\left(\Xi_2,B_{\Sigma}B_{\del M}\,\Psi_2\right) = \frac{\alpha'}{2}\w\left(\Xi_2,\delta_\Sigma\Omega_\Sigma c_0^+(-\partial_\varphi^*\delta_{\partial M}+\delta_{\partial M}\partial_\varphi)\,\Psi_2\right)\,.
\end{equation}
Next, note that all the terms with $\varphi$ Virasoro generators in \ref{Ghost number zero state} will vanish in the overlap, since $\delta_\Sigma\Omega_\Sigma c_0^+(\partial_\varphi^*\delta+\delta\partial_\varphi)\,\Psi_2$ contains no $\varphi$ oscillators. Thus, only four terms in \ref{Ghost number zero state} contribute to the charge. Let us consider the first term in \ref{Xi_2}. The holomorphic part gives
\begin{subequations}
\begin{align}
    &i\left(\frac{\alpha'}{2}\right)^{3/2}\omega_L\left(\Xi^{(0)}_{2L},\delta_\Sigma\left(\sum_{n\neq0}c_{-n}\alpha_n^t\right) c_0^+(-\partial_\varphi^*\delta_{\partial M}+\delta_{\partial M}\partial_\varphi)\lambda_2c_1P_2e^{3V\varphi/2}|0\rangle\right) =\nonumber \\ 
    &\hspace{1cm}= i\left(\frac{\alpha'}{2}\right)^{3/2}\w_L\left(e^{-V\varphi/2},\delta_\Sigma c_0^+(-\partial_\varphi^*\delta_{\partial M}+\delta_{\partial M}\partial_\varphi)\,(4b_{3}c_{-1}\alpha^t_1c_{-3}\alpha^t_3+ \right.\nonumber\\
   &\left.\hspace{2cm} - 2b_2c_{-1}\alpha^t_2c_{-2}\alpha^t_2 + (\alpha^{t\,3}_1+2\alpha_3)c_{-1}\alpha_1-2\alpha_3c_{-1}\alpha_1)\lambda_2c_1P_2e^{3V\varphi/2}|0\rangle\right)\\ 
   &\hspace{1cm}= -i3\sqrt{3}\alpha'^{3/2}V\lambda_2\,,
\end{align}
\end{subequations}
where $\w_L$ includes the contributions of the holomorphic oscillators and of the $\varphi$ zero mode. The antiholomorphic factor is simply $1$. Finally, the second term in \ref{Xi_2} gives an identical contribution. All in all, the result is
\begin{align}
    \w\left(\Xi_2,B_{\Sigma}B_{\del M}\,\Psi_2\right)
    = &3\sqrt{6}\alpha'V\lambda_2\,.
\end{align}
Although we did not check this explicitly, it is plausible that the construction in \cite{Imbimbo:1991ia} will generate the zero-momentum ghost number zero cohomology for higher $r$, and thus the higher level isometries  to compute all the black hole charges $\lambda_r$.

\section{Details of the interacting charge 1-form} \label{app:B}

In this Appendix, we will fill in the technical details concerning the definition and properties of the conserved charge for interacting SFTs presented in Section \ref{sec:interact}. 

\subsection{Consistency of $\Xi_\Psi$}

Let us start by showing that the expression
\begin{align}
    E_\Psi = -h_\Psi l_2^\Psi(\delta\Psi,\Xi_\Psi)
\end{align}
is a closed 1-form so that (at least locally) it can be integrated between two classical field configurations independently of the choice of the path to give the iterative expression \eqref{eq:Xiimplicit} for $\Xi_\Psi$. In computing $\delta E_\Psi$, we can first make use of the fact that the shifted $L_\infty$ products satisfy
\begin{subequations}
    \begin{align}
        \delta Q_\Psi \cdot  &=-l_2^\Psi(\delta \Psi,\cdot)\,,\\
        \delta l_2^\Psi( \cdot,\cdot)  &=-l_3^\Psi(\delta \Psi,\cdot,\cdot)\,.\\
         &\hspace{2mm}\vdots\nonumber
    \end{align}
\end{subequations}
Furthermore, taking into account the the Hodge-Kodaira decomposition \eqref{eq:HK}, together with the conditions \eqref{eq:PQ} and \eqref{eq:side}, we can find that
\begin{subequations}
    \begin{align}
        \delta h_\Psi\cdot &= -h_\Psi l_2^\Psi(\delta\Psi,h_\Psi\cdot )\,,\\
        \delta P_\Psi\cdot &= P_\Psi l_2^\Psi(\delta\Psi,\cdot)h_\Psi-h_\Psi l_2^\Psi(\delta\Psi,\cdot)P_\Psi\,.
    \end{align}
\end{subequations}
Finally, for $\delta\Xi_\Psi$ we substitute $E_\Psi$ again.
Altogether, this gives
\begin{subequations}
    \begin{align}
    \delta E_\Psi &= -\delta h_\Psi l_2^\Psi(\delta\Psi,\Xi_\Psi) +h_\Psi \delta l_2^\Psi(\delta\Psi,\Xi_\Psi)+h_\Psi l_2^\Psi(\delta\Psi,\delta\Xi_\Psi)\\
    &= h_\Psi l_2^\Psi(\delta\Psi,h_\Psi l_2^\Psi(\delta\Psi,\Xi_\Psi) )  -h_\Psi l_3^\Psi(\delta \Psi,\delta\Psi,\Xi_\Psi) -h_\Psi l_2^\Psi(\delta\Psi,h_\Psi l_2^\Psi(\delta\Psi,\Xi_\Psi))\\
    &=0\,,
\end{align}
\end{subequations}
where the last equality holds because the first and the third term cancel, while the second term is zero by virtue of the graded-symmetry property of $l_3^\Psi$. Hence, the 1-form $E_\Psi$ is closed and therefore locally exact, as claimed.

\subsection{Closedness of the charge 1-form}

We will proceed by proving the closedness of the charge 1-form $\eta(\Xi_\Psi,\Psi)$. Realizing that the differentials of the boundary operators can be computed as
\begin{subequations}
    \begin{align}
        \delta B_{\p M}^\Psi\cdot &= -[\Theta_M,l_2^\Psi(\delta \Psi,\cdot)]\,,\\
        \delta B_{\Sigma}^\Psi\cdot &= -[\Theta_-,l_2^\Psi(\delta \Psi,\cdot)]\,,
    \end{align}
\end{subequations}
we can differentiate $\eta(\Xi_\Psi,\Psi)$ to obtain (recalling that both $\delta$ and $\omega$ are anti-commuting objects and reversing the order of the boundary operators inside $\eta(\Xi_\Psi,\Psi)$ for the sake of later convenience)
\begin{align}
  \delta\eta(\Xi_\Psi,\Psi)&=  \underbrace{\omega(\delta\Xi_\Psi,  B^{\Psi}_{\partial M} B^{\Psi}_\Sigma\delta \Psi)}_{\equiv K }+\nonumber\\
  &\hspace{2cm}\underbrace{+\omega(\Xi_\Psi, [\Theta_M,l_2^\Psi(\delta \Psi,\cdot)][\Theta_-,Q_\Psi] \delta \Psi)}_{\equiv L_1}+\nonumber\\
  &\hspace{2cm}\underbrace{-\omega(\Xi_\Psi, [\Theta_M,Q_\Psi][\Theta_-,l_2^\Psi(\delta \Psi,\cdot)]  \delta \Psi)}_{\equiv L_2}\,.
\end{align}
We have organized the computation as a sum of three terms $\delta \eta (\Xi_\Psi,\Psi) = K+L_1+L_2$. Starting by manipulating $L_1$, we can open up the commutator of $Q_\Psi$ with $\Theta_-$ to write
\begin{subequations}
   \begin{align}
    L_1&=-\omega(\Xi_\Psi, [\Theta_M,l_2^\Psi(\delta \Psi,\cdot)]Q_\Psi\Theta_- \delta \Psi)\\
    &=-\omega\Big(\Xi_\Psi, Q_\Psi[\Theta_M,l_2^\Psi(\delta \Psi,\cdot)]\Theta_- \delta \Psi-[[Q_\Psi,\Theta_M],l_2^\Psi(\delta \Psi,\cdot)]\Theta_- \delta \Psi\Big)\\
    &=-\omega\Big(\Xi_\Psi, Q_\Psi\big([\Theta_M,l_2^\Psi(\delta \Psi,\cdot)]\Theta_- \delta \Psi-[\Theta_M,l_2^\Psi(\Theta_-\delta \Psi,\cdot)] \delta \Psi\big)\Big)+\nonumber\\
    &\hspace{2cm}    
    -\omega\Big(\Xi_\Psi,Q_\Psi[\Theta_M,l_2^\Psi(\Theta_-\delta \Psi,\cdot)] \delta \Psi-[[Q_\Psi,\Theta_M],l_2^\Psi(\delta \Psi,\cdot)]\Theta_- \delta \Psi\Big)\,.\label{eq:3}
\end{align} 
\end{subequations}
In the third equality, we notice that the terms on the first line are localized on $\p M\cap \Sigma$: if $\Theta_-$ were a constant, the first line would have vanished, meaning that it is proportional to a derivative of $\Theta_-$, thus localizing the expression on $\Sigma$. Moreover, $\Theta_M$ only enters through commutators so that an analogous argument applies. This means that in the first line of \eqref{eq:3}, $Q_\Psi$ can be integrated by parts with impunity to kill $\Xi_\Psi$ on the bra. We can therefore see that by this manipulation, we have effectively managed to move the insertion of $\Theta_-$ around the expression, ending up with
\begin{subequations}
\label{eq:L12}
   \begin{align}
    L_1&=- \omega\Big(\Xi_\Psi,[[Q_\Psi,\Theta_M],l_2^\Psi(\Theta_-\delta \Psi,\cdot)] \delta \Psi-
    [\Theta_M,l_2^\Psi([Q_\Psi,\Theta_-]\delta \Psi,\cdot)] \delta \Psi+\nonumber\\
    &\hspace{8cm}
    -[[Q_\Psi,\Theta_M],l_2^\Psi(\delta \Psi,\cdot)]\Theta_- \delta \Psi\Big)\\
    &= -\omega\Big(\Xi_\Psi,-2[\Theta_M,Q_\Psi][\Theta_-,l_2^\Psi(\delta \Psi,\cdot)]\delta \Psi +[\Theta_M,l_2^\Psi(\delta \Psi,\cdot)][\Theta_-,Q_\Psi]\delta \Psi 
    +\nonumber\\
    &\hspace{3cm}
    -l_2^\Psi(\delta \Psi,[\Theta_M,Q_\Psi]\Theta_- \delta \Psi)+l_2^\Psi(\Theta_-\delta \Psi,[\Theta_M,Q_\Psi] \delta \Psi)  +\nonumber\\
    &\hspace{3cm}
      -l_2^\Psi(\Theta_M \delta \Psi,[\Theta_-,Q_\Psi]\delta \Psi)
    +l_2^\Psi( \delta \Psi,\Theta_M[\Theta_-,Q_\Psi]\delta \Psi)
    \Big)  \,,\label{eq:L12b}
\end{align} 
\end{subequations}
where purely algebraic manipulations were performed when going from the first line of \eqref{eq:L12} to the second with no integration by parts performed on individual terms. At each step of the calculation, the overall spacetime integral implied by the symplectic pairing localizes on $\p M\cap \Sigma$ so that, in particular, one can assume that $\Psi$ satisfies the equation of motion \eqref{eq:Linfty}. Moving the terms on the first line of \eqref{eq:L12b} to the l.h.s. and dividing by 2, we therefore obtain the result
\begin{align}
    L_1+L_2 &= -\frac{1}{2} \omega\Big(\Xi_\Psi,
    l_2^\Psi(\Theta_-\delta \Psi,[\Theta_M,Q_\Psi] \delta \Psi)-l_2^\Psi(\delta \Psi,[\Theta_M,Q_\Psi]\Theta_- \delta \Psi)\Big)  +\nonumber\\
    &\hspace{2cm}
      +\frac{1}{2} \omega\Big(\Xi_\Psi,l_2^\Psi(\Theta_M \delta \Psi,[\Theta_-,Q_\Psi]\delta \Psi)
      - l_2^\Psi( \delta \Psi,\Theta_M[\Theta_-,Q_\Psi]\delta \Psi)
    \Big)\,.
\end{align}
Both terms separately (and manifestly) localize on $\p M\cap \Sigma$. Continuing with $K$, we can first use cyclicity on $\p M\cap \Sigma$ to rewrite
\begin{align}
    K=-\omega( \Xi_\Psi,l_2^\Psi(\delta\Psi,  h_\Psi B^{\Psi}_{\partial M} B^{\Psi}_\Sigma\delta \Psi))\,.
\end{align}
Furthermore, opening up the commutator in the boundary operator $B^{\Psi}_\Sigma$, we can write
\begin{subequations}
    \begin{align}
        K&=\omega( \Xi_\Psi,l_2^\Psi(\delta\Psi,  h_\Psi [\Theta_M,Q_\Psi] Q_\Psi\Theta_- \delta\Psi))\\
        &=\omega\Big( \Xi_\Psi,-l_2^\Psi(\delta\Psi,  [h_\Psi, Q_\Psi][\Theta_M,Q_\Psi] \Theta_- \delta\Psi)+Q_\Psi l_2^\Psi(\delta\Psi,  h_\Psi [\Theta_M,Q_\Psi] \Theta_- \delta\Psi)\Big)\,.\label{eq:Kb}
    \end{align}
\end{subequations}
Again, we will be making use of the fact the as long as we do not break up the overall spacetime integral contained in $\omega$ to individual terms (which may not themselves separately localize on $\p M\cap \Sigma$), we can assume that the position zero-mode in all terms of the integrand localizes on $\p M\cap \Sigma$ and therefore that the equation of motion \eqref{eq:Linfty} is satisfied.
In the first term in \eqref{eq:Kb}, we can substitute from the Hodge-Kodaira decomposition \eqref{eq:HK} while in the second term, a similar reasoning as in \eqref{eq:3} can be used to move around the insertion of $\Theta_-$: in particular, we notice that the expression
\begin{align}
    Q_\Psi l_2^\Psi(\delta\Psi,  h_\Psi [\Theta_M,Q_\Psi] \Theta_- \delta\Psi)-Q_\Psi l_2^\Psi(\Theta_- \delta\Psi,  h_\Psi [\Theta_M,Q_\Psi] \delta\Psi)
\end{align}
localizes on the intersection $\p M\cap\Sigma$, so that the overall $Q_\Psi$ action can be brought on the bra of the symplectic pairng where it kills $\Xi_\Psi$. This allows us to rewrite
\begin{subequations}
\begin{align}
    K &=\omega\Big( \Xi_\Psi,-l_2^\Psi(\delta\Psi,  (1-P_\Psi)[\Theta_M,Q_\Psi] \Theta_- \delta\Psi)+Q_\Psi l_2^\Psi(\Theta_-\delta\Psi,  h_\Psi [\Theta_M,Q_\Psi]  \delta\Psi)\Big)\\
    &=\omega\Big( \Xi_\Psi,-l_2^\Psi(\delta\Psi,  (1-P_\Psi)[\Theta_M,Q_\Psi] \Theta_- \delta\Psi)+ l_2^\Psi([\Theta_-,Q_\Psi]\delta\Psi,  h_\Psi [\Theta_M,Q_\Psi]  \delta\Psi)+\nonumber\\
    &\hspace{7.2cm}+ l_2^\Psi(\Theta_-\delta\Psi,  (1-P_\Psi)[\Theta_M,Q_\Psi]  \delta\Psi)
    \Big)\,.\label{eq:Kb2}
\end{align}
\end{subequations}
In the last term of \eqref{eq:Kb2}, we notice that only the identity contribution from the Hodge-Kodaira decomposition survives because
\begin{align}
    P_\Psi[\Theta_M,Q_\Psi]\delta\Psi = P_\Psi\Theta_MQ_\Psi\delta\Psi-P_\Psi Q_\Psi \Theta_M\delta\Psi=0\,,\label{eq:PTQM}
\end{align}
where the first term is zero courtesy of the equations of motion while the second terms vanishes due to the condition \eqref{eq:PQ} on the projector $P_\Psi$, which we take to project on the cohomology of $Q_\Psi$. To manipulate the second term in \eqref{eq:Kb2}, we recall that $h_\Psi$ projects \textit{out} of the $Q_\Psi$-cohomology, so that $h_\Psi\delta\Psi=0$ and we can rewrite
\begin{subequations}
\begin{align}
    h_\Psi [\Theta_M,Q_\Psi]  \delta\Psi &= [h_\Psi, [\Theta_M,Q_\Psi] ] \delta\Psi \\
    &= [[h_\Psi,\Theta_M],Q_\Psi]  \delta\Psi+ [\Theta_M,[h_\Psi,Q_\Psi]]  \delta\Psi\label{eq:hTQb}\\
    &= [[h_\Psi,\Theta_M],Q_\Psi]  \delta\Psi+ [\Theta_M,-P_\Psi]  \delta\Psi\,,\label{eq:hTQc}
\end{align}
\end{subequations}
where the first term in \eqref{eq:hTQc} will not contribute upon inserting it back in the second term in \eqref{eq:Kb2} because the commutators $[h_\Psi,\Theta_M]$ and $[\Theta_-,Q_\Psi]$ localize the expression on $\p M\cap \Sigma$ so that the commutator $[[h_\Psi,\Theta_M],Q_\Psi]$ can be opened up and integration by parts then gives zero. Finally, we can notice that the projector part of the first term in \eqref{eq:Kb2} can be rewritten as
\begin{subequations}
    \begin{align}
        l_2^\Psi(\delta\Psi,  P_\Psi[\Theta_M,Q_\Psi] \Theta_- \delta\Psi)&=l_2^\Psi(\delta\Psi,  P_\Psi\Theta_MQ_\Psi \Theta_- \delta\Psi)\\
        &=l_2^\Psi(\delta\Psi,  P_\Psi\Theta_M[Q_\Psi, \Theta_-] \delta\Psi)\\
        &=l_2^\Psi(\delta\Psi,  [P_\Psi,\Theta_M][Q_\Psi, \Theta_-] \delta\Psi)\,,
    \end{align}
\end{subequations}
where in the last equality we have used that $P_\Psi[Q_\Psi,\Theta_-]\delta\Psi=0$, in an analogy with the result \eqref{eq:PTQM}.
Altogether, we therefore obtain
\begin{align}
    K&=+\omega\Big( \Xi_\Psi, l_2^\Psi(\Theta_-\delta\Psi,  [\Theta_M,Q_\Psi]  \delta\Psi)-l_2^\Psi(\delta\Psi,  [\Theta_M,Q_\Psi] \Theta_- \delta\Psi)\Big)+\nonumber\\
    &\hspace{0.2cm}
    +\omega\Big( \Xi_\Psi,l_2^\Psi(\delta\Psi,  [\Theta_M,P_\Psi][\Theta_-,Q_\Psi] \delta\Psi)\Big)-\omega\Big(\Xi_\Psi, l_2^\Psi( [\Theta_M,P_\Psi]  \delta\Psi,[\Theta_-,Q_\Psi]\delta\Psi)
    \Big)\,,
\end{align}
where all three integrands on the first and the second line manifestly localize $\omega$ on the intersection $\p M\cap \Sigma$.
At the same time, instead of manipulating $K$ by opening up the commutator in the boundary operator $B_\Sigma^\Psi$, we could have opened up the commutator in $B_{\p M}^\Psi$. Following analogous steps, we would have obtained
\begin{align}
    K&=-\omega\Big( \Xi_\Psi, l_2^\Psi(\Theta_M\delta\Psi,  [\Theta_-,Q_\Psi]  \delta\Psi)-l_2^\Psi(\delta\Psi,  [\Theta_-,Q_\Psi] \Theta_+ \delta\Psi)\Big)+\nonumber\\
    &\hspace{0.2cm}
    -\omega\Big( \Xi_\Psi,l_2^\Psi(\delta\Psi,  [\Theta_-,P_\Psi][\Theta_M,Q_\Psi] \delta\Psi)\Big)+\omega\Big(\Xi_\Psi, l_2^\Psi( [\Theta_-,P_\Psi]  \delta\Psi,[\Theta_M,Q_\Psi]\delta\Psi)
    \Big)\,.
\end{align}
Averaging the two expressions, we can therefore write
    \begin{align}
     K&=   \frac{1}{2}\omega\Big( \Xi_\Psi, l_2^\Psi(\Theta_-\delta\Psi,  [\Theta_M,Q_\Psi]  \delta\Psi)-l_2^\Psi(\delta\Psi,  [\Theta_M,Q_\Psi] \Theta_- \delta\Psi)\Big)+\nonumber\\
     &\hspace{3cm}-\frac{1}{2}\omega\Big( \Xi_\Psi, l_2^\Psi(\Theta_M\delta\Psi,  [\Theta_-,Q_\Psi]  \delta\Psi)-l_2^\Psi(\delta\Psi,  [\Theta_-,Q_\Psi] \Theta_+ \delta\Psi)\Big)+\nonumber\\
    &\hspace{-0.5cm}
    +\frac{1}{2}\omega\Big( \Xi_\Psi,l_2^\Psi(\delta\Psi,  [\Theta_M,P_\Psi][\Theta_-,Q_\Psi] \delta\Psi)\Big)-\frac{1}{2}\omega\Big(\Xi_\Psi, l_2^\Psi( [\Theta_M,P_\Psi]  \delta\Psi,[\Theta_-,Q_\Psi]\delta\Psi)
    \Big)+\nonumber\\
     &\hspace{-0.5cm}
    -\frac{1}{2}\omega\Big( \Xi_\Psi,l_2^\Psi(\delta\Psi,  [\Theta_-,P_\Psi][\Theta_M,Q_\Psi] \delta\Psi)\Big)+\frac{1}{2}\omega\Big(\Xi_\Psi, l_2^\Psi( [\Theta_-,P_\Psi]  \delta\Psi,[\Theta_M,Q_\Psi]\delta\Psi)
    \Big)\,,
    \end{align}
where in the first two terms, we recognize precisely $-L_1-L_2$. This allows us to rewrite the overall differential of the 1-form $\eta(\Xi_\Psi,\Psi)$ as
\begin{subequations}
\begin{align}
     &\delta\eta(\Xi_\Psi,\Psi)=\nonumber\\
     &\hspace{-0.3cm}=K+L_1+L_2\\
     &\hspace{-0.3cm}=\frac{1}{2}\omega\Big( \Xi_\Psi,l_2^\Psi(\delta\Psi,  [\Theta_M,P_\Psi][\Theta_-,Q_\Psi] \delta\Psi)\Big)   -\frac{1}{2}\omega\Big( \Xi_\Psi,l_2^\Psi(\delta\Psi,  [\Theta_-,P_\Psi][\Theta_M,Q_\Psi] \delta\Psi)\Big)+\nonumber\\
     &\hspace{-0.2cm}-\frac{1}{2}\omega\Big(\Xi_\Psi, l_2^\Psi( [\Theta_M,P_\Psi]  \delta\Psi,[\Theta_-,Q_\Psi]\delta\Psi)
    \Big)
 +\frac{1}{2}\omega\Big(\Xi_\Psi, l_2^\Psi( [\Theta_-,P_\Psi]  \delta\Psi,[\Theta_M,Q_\Psi]\delta\Psi)
    \Big)\,.\label{eq:deta}
\end{align}
\end{subequations}
To address the first two terms in \eqref{eq:deta}, we can make use of the possibility to freely integrate by parts to apply cyclicity of $l_2^\Psi$. One obtains, e.g.\ for the first term,
\begin{subequations}
\begin{align}
    \omega\Big( \Xi_\Psi,l_2^\Psi(\delta\Psi,  [\Theta_M,P_\Psi][\Theta_-,Q_\Psi] \delta\Psi)\Big)   &=    \omega\Big( l_2^\Psi(\Xi_\Psi,\delta\Psi),  [\Theta_M,P_\Psi][\Theta_-,Q_\Psi] \delta\Psi\Big)\\
    &=\omega\Big( l_2^\Psi(\Xi_\Psi,\delta\Psi),P_\Psi  [\Theta_M,P_\Psi][\Theta_-,Q_\Psi] \delta\Psi\Big)\,,\label{eq:finalb}
\end{align}
\end{subequations}
where to write the second equality, we recalled the step from \eqref{eq:hTQb} to \eqref{eq:hTQb}, as well as the fact that $P_\Psi^2=P_\Psi$. By now, however, it is manifest that the r.h.s.\ of \eqref{eq:finalb} has to vanish, because the projection of $l_2^\Psi(\Xi_\Psi,\delta\Psi)$ onto the cohomology needs to be zero in order for the isometry $\Xi_\Psi$ to be well defined on the phase space (which amounts to requiring that there is no obstruction to solving \eqref{eq:delXi} -- see the discussion in Section \ref{sec:interact} for more details). The second term in \eqref{eq:deta} can be dealt with similarly.

Finally, we have to show that the third and fourth term in \eqref{eq:deta} also vanish. Starting e.g. with the third term, we can write
\begin{subequations}
    \begin{align}
       &-\frac{1}{2}\omega\Big(\Xi_\Psi, l_2^\Psi( [\Theta_M,P_\Psi]  \delta\Psi,[\Theta_-,Q_\Psi]\delta\Psi)
    \Big)=\nonumber\\
    &\hspace{1cm}=\frac{1}{2}\omega\Big(\Xi_\Psi, l_2^\Psi( [\Theta_M,P_\Psi]  \delta\Psi,Q_\Psi\Theta_-\delta\Psi)
    \Big)\\
    &\hspace{1cm}=\frac{1}{2}\omega\Big(\Xi_\Psi,Q_\Psi l_2^\Psi( [\Theta_M,P_\Psi]  \delta\Psi,\Theta_-\delta\Psi)+ l_2^\Psi( Q_\Psi[\Theta_M,P_\Psi]  \delta\Psi,\Theta_-\delta\Psi)
    \Big)\,.\label{eq:22b}
    \end{align}
\end{subequations}
At this point, we notice that since $\delta\Psi$ is in the $Q_\Psi$-cohomology, we have $\delta\Psi=P_\Psi\delta\Psi$ so that the first term in the integrand of \eqref{eq:22b} can be rewritten as
\begin{subequations}
    \begin{align}
        &Q_\Psi l_2^\Psi( [\Theta_M,P_\Psi]  \delta\Psi,\Theta_-\delta\Psi)=\nonumber\\
        &\hspace{1cm}=Q_\Psi l_2^\Psi( [\Theta_M,P_\Psi]  \delta\Psi,\Theta_-P_\Psi\delta\Psi)\\
        &\hspace{1cm}=Q_\Psi l_2^\Psi( [\Theta_M,P_\Psi]  \delta\Psi,P_\Psi\Theta_-\delta\Psi)+Q_\Psi l_2^\Psi( [\Theta_M,P_\Psi]  \delta\Psi,[\Theta_-,P_\Psi]\delta\Psi)\,,\label{eq:23b}
    \end{align}
\end{subequations}
where the second term in \eqref{eq:23b} can be dropped because it is localized on $\p M\cap \Sigma$, so the overall $Q_\Psi$ can be integrated by parts to hit $\Xi_\Psi$ on the bra which, in turn, gives 0. Bringing the $Q_\Psi$ back inside $l_2^\Psi$ in the first term in \eqref{eq:23b}, recalling that $QP_\Psi=0$ and substituting back into \eqref{eq:22b}, we therefore obtain
\begin{subequations}
    \begin{align}
       &-\frac{1}{2}\omega\Big(\Xi_\Psi, l_2^\Psi( [\Theta_M,P_\Psi]  \delta\Psi,[\Theta_-,Q_\Psi]\delta\Psi)
    \Big)=\nonumber\\
    &\hspace{1cm}=\frac{1}{2}\omega\Big(\Xi_\Psi, -l_2^\Psi( Q_\Psi[\Theta_M,P_\Psi]  \delta\Psi,P_\Psi\Theta_-\delta\Psi)+ l_2^\Psi( Q_\Psi[\Theta_M,P_\Psi]  \delta\Psi,\Theta_- P_\Psi\delta\Psi)
    \Big)\\
      &\hspace{1cm}=\frac{1}{2}\omega\Big(\Xi_\Psi, l_2^\Psi( Q_\Psi[\Theta_M,P_\Psi]  \delta\Psi,[\Theta_-, P_\Psi]\delta\Psi)
    \Big)\\
      &\hspace{1cm}=-\frac{1}{2}\omega\Big(\Xi_\Psi, l_2^\Psi([\Theta_-, P_\Psi]\delta\Psi, [\Theta_M,Q_\Psi]  \delta\Psi)
    \Big)\,,
    \end{align}
\end{subequations}
which is exactly minus the fourth term in \eqref{eq:deta}. This shows that the third and fourth term in \eqref{eq:deta} exactly cancel, thus concluding the proof that
\begin{align}
    \delta \eta(\Xi_\Psi,\Psi) =0\,.
\end{align}


\endgroup

\end{document}